\documentclass[usenatbib]{mn2e}

\usepackage{apjfonts,amsfonts,verbatim}
\usepackage{bm}
\usepackage{mathtools}
\usepackage{xcolor}
\usepackage{natbib}
\usepackage{lscape}
\usepackage{afterpage}  
\usepackage{booktabs}
\usepackage{tabularx}
\usepackage{adjustbox}

\usepackage{threeparttable}

\usepackage[T1]{fontenc}

\usepackage{graphicx}	
\usepackage{amsmath}	
\usepackage{amssymb}	

\newcommand{\msun}{\ensuremath{M_\odot}}
\newcommand{\lya}{Ly$\alpha$}

\newcommand{\oi}{\rm O\,{\textsc {i}}}
\newcommand{\HI}{\rm H\,{\textsc {i}}}

\newcommand{\SiII}{\rm Si\,{\textsc {ii}}}
\newcommand{\CII}{\rm C\,{\textsc {ii}}}
\newcommand{\MgII}{\rm Mg\,{\textsc {ii}}}

\newcommand{\kms}{\,\ifmmode{\mathrm{km}\,\mathrm{s}^{-1}}\else km\,s${}^{-1}$\fi}

\title[Modeling Metal Absorption Lines]{\texttt{ALPACA}: A New Semi-Analytic Model for Metal Absorption Lines Emerging from Clumpy Galactic Environments}

\author[Li et al.]{Zhihui Li,$^{1}$\thanks{E-mail: zhihui@caltech.edu}
Max Gronke$^{2}$, and Charles C. Steidel$^{1}$
\\
$^{1}$Cahill Center for Astronomy and Astrophysics, California Institute of Technology, 1200 E California Blvd, MC 249-17, Pasadena, CA 91125, USA\\
$^{2}$Max-Planck Institute for Astrophysics, Karl-Schwarzschild-Str. 1, D-85741 Garching, Germany
}

\date{}

\begin{document}
\label{firstpage}
\pagerange{\pageref{firstpage}--\pageref{lastpage}}
\maketitle

\begin{abstract}
We present a new semi-analytic formalism for modeling metal absorption lines that emerge from a clumpy galactic environment, \texttt{ALPACA}. We predict the ``down-the-barrel'' (DTB) metal absorption line profiles and the EW of absorption at different impact parameters as a function of the properties of the clumps, including the clump kinematics, the clump volume filling factor, the clump number density profile and the clump ion column densities. With \texttt{ALPACA}, we jointly model the stacked DTB \CII\,$\lambda$1334 spectrum of a sample of $z \sim$ 3 Lyman break galaxies and the EW v.s. $b$ profile of a sample of $z \sim$ 2 star-forming galaxy-galaxy pairs. \texttt{ALPACA} successfully reproduced two datasets simultaneously, and the best-fit prefers a low clump volume filling factor ($\sim 3 \times 10^{-3}$). The radial velocities of the clumps are a superposition of a rapidly accelerated outflow with a maximum velocity of $\sim 400 \kms$ and a velocity dispersion of $\sigma_{\rm cl} \sim 120 \kms$. The joint modeling reveals a physical scenario where the absorption observed at a particular velocity is contributed by the clumps distributed over a fairly broad range of radii. We also find that the commonly adopted Sobolev approximation is at best only applicable within a narrow range of radii where the clumps are undergoing rapid acceleration in a non-volume-filling clumpy medium. Lastly, we find that the clump radial velocity profile may not be fully constrained by the joint modeling and spatially-resolved \lya\ emission modeling may help break the degeneracy.

\end{abstract}


\section{Introduction}

Metal absorption lines observed in the rest-frame ultraviolet (UV) encode abundant information about the physical properties of the gaseous matter in a galactic environment -- from the interstellar medium (ISM; \citealt{Tacconi20}) to the circumgalactic medium (CGM; \citealt{Tumlinson17,FG2023}) to the intergalactic medium (IGM; \citealt{McQuinn16}). Such absorption lines are typically produced via the transition of an atom or ion from the ground state to an excited state by absorbing the energetic UV continuum photons produced in star-forming regions. Depending on whether the ground state is further split into fine-structure levels, such transitions can be either resonant (e.g. \lya\ and \MgII\,$\lambda\lambda$2796, 2803) or non-resonant (e.g. \SiII\,$\lambda$1260 and \CII\,$\lambda$1334), the latter of which is considered to have ``fluorescent'' channels through which the photons at the resonant wavelength can be emitted at a slightly lower energy.

A typical metal absorption line observed against a galaxy's own starlight (namely ``down-the-barrel"; DTB) is ``sawtooth'' shaped (e.g. \citealt{Weiner09,Rubin10, Martin12}), although in reality it exhibits a wide variety of spectral morphologies. Specifically, the minimum flux density (the ``trough'') is often located at a few hundred \kms\ blueward (or even redward in rare cases; see e.g. \citealt{Rubin12,Martin12,Bouche13,Ford14,Ho17,Zabl19,Afruni22,Weldon23}) of the systemic velocity. On both sides of the trough, the flux density gradually rises to meet the continuum, yet in general, it rises significantly more steeply on the red side than the blue side. The spectral features of the metal absorption lines can then be used to infer the physical properties of the absorbing gas. For example, the velocity range of the absorption line profile traces the gas outflow velocities, and the depth of the absorption probes the gas column density or covering fraction. In particular, the absorption lines from low-ionization states (LIS), such as \SiII, \CII\ and \oi, closely trace neutral hydrogen due to their similar ionization potential. The derived gas properties from the LIS lines can therefore be utilized to constrain several important galactic properties, such as the mass outflow rates, the escape fraction of ionizing photons, etc. (e.g., \citealt{Rupke05, Martin05, Weiner09, Martin09, Rubin14, Erb15, Chisholm16, Chisholm18, Steidel18, Gazagnes18, Gazagnes20, Mauerhofer21, Xu22}).

Thus far, a number of attempts have been made to model the metal absorption lines in DTB galaxy spectra. Most have adopted a ``picket-fence'' model (e.g. \citealt{Steidel10, Heckman11, Zackrisson13, Jones13,  Borthakur14, Rivera15, Reddy16, Rivera2017, Steidel18, Gazagnes18, Gazagnes20, Xu22}), which assumes that the stellar continuum is partially covered by optically thick absorbing gaseous material. Some have further accounted for radial variation of the gas outflow velocity to reproduce the line profiles of  particular transitions (e.g. \citealt{Steidel10, Chisholm16}). Other work has explored, using semi-analytic models or Monte Carlo simulations, the absorption line profile resulting from transmission through a homogeneous, expanding wind  (e.g. \citealt{Prochaska11,Scarlata15, Carr21}, while others have used cosmological simulations to predict the absorption line profiles emerging from realistic galactic environments (e.g. \citealt{Kimm11, Mauerhofer21, Gazagnes23}). Many of the models have successfully produced absorption line profiles that closely resemble observations. Nevertheless, the majority of the models proposed in previous works rely on simplifying assumptions, e.g. that the gas column density is always high enough to result in saturated absorption so that the depth of absorption relative to the continuum directly traces the gas covering fraction; that continuum photons will be absorbed by the outflowing gas with a large velocity gradient only if they appear resonant in the reference frame of the gas (namely the Sobolev approximation), or that the absorbing gaseous medium is homogeneous without any holes or clumps. These assumptions may be (at least in part) unphysical or in tension with the most recent observations. For example, theoretical models, simulations and observations have revealed that galactic winds may reach a ``plateau'' phase at large radii where the wind velocity remains approximately constant (e.g. \citealt{Chevalier85,Veilleux05,Dorfi12,Zhang18}). Recent work also highlighted the importance of accounting for the multiphase, turbulent and kinematically complex structure of galactic winds \citep{Schneider20,Kim20,Fielding22,Steinwandel22b,Steinwandel22a,Rathjen23}. As these recent findings have posed significant challenges to the aforementioned simplifying assumptions, the models that depend on them should benefit from re-examination.

In this work, we build on previous models and present a new semi-analytic model for the UV metal absorption lines. Thus far, the clumpy nature of the ``cool'' ($T \sim 10^4 \rm$\,K) gas in the ISM / CGM has been supported by abundant observational evidence (e.g. \citealt{Rauch99,Rauch01,Rauch01b,Rauch02,Ellison04,Schaye07,Rogerson12,Crighton15,Arrigoni15,Rubin18,Kulkarni19,Zahedy19,Zahedy21}). More specifically, the cool gas (which is responsible for producing the LIS lines) is likely to exist in the form of a clumpy mist or fog of cloudlets with a large area covering fraction but a small volume filling factor \citep{McCourt18,Fielding20,Gronke20,Nelson20}. In light of this physical picture, we explore the formation of metal absorption lines from a clumpy galactic outflow. Not only do we model the DTB absorption line profiles, but we also predict the strength of absorption as a function of impact parameter \citep{Steidel10}. The ultimate goal of this work is to develop a simple, usable model for the community to fit and interpret the observed metal absorption lines fast and robustly. 

The structure of this paper is as follows. In \S\ref{sec:clumpy_model}, we describe the general formalism and a practical implementation of the analytic model. In \S\ref{sec:valid}, we validate the analytic model by comparing it to Monte-Carlo numerical simulations. In \S\ref{sec:params}, we discuss the effect of each individual parameter of the analytic model. In \S\ref{sec:data}, we show an example of applying the analytic model to the composite \CII\,$\lambda$1334 spectrum of a sample of $z \sim 3$ Lyman break galaxies (LBG) observed for the Keck Lyman Continuum Spectroscopic Survey (KLCS; \citealt{Steidel18}) and the EW v.s. $b$ profile observed for a sample of $z \sim 2$ star-forming galaxy-galaxy pairs. In \S\ref{sec:fc_Fv}, we discuss the definition and relationship between the gas covering and volume filling parameters. In \S\ref{sec:non_sobolev}, we compare the models that use or not use the Sobolev approximation. In \S\ref{sec:previous_works}, we discuss previous work modeling the UV absorption lines in comparison with our model. In \S\ref{sec:caveats_outlook}, we discuss the limitations of our model and possible developments in the future. In \S\ref{sec:conclusion}, we summarize and conclude. 

\section{\texttt{ALPACA}: A Non-Sobolev Clumpy Model For Metal Absorption Lines}\label{sec:clumpy_model}
We introduce the semi-analytic model that we use in this work, \texttt{ALPACA} (Absorption Line Profiles Arising from Clumpy Absorbers)\footnote{The code for \texttt{ALPACA} is publicly available at: \url{https://github.com/astro-zhihuili/ALPACA}.}.

\subsection{General Formalism}\label{sec:formalism}
\subsubsection{Down the Barrel Absorption}\label{sec:dtb}
\begin{figure}
\centering
\includegraphics[width=0.47\textwidth]{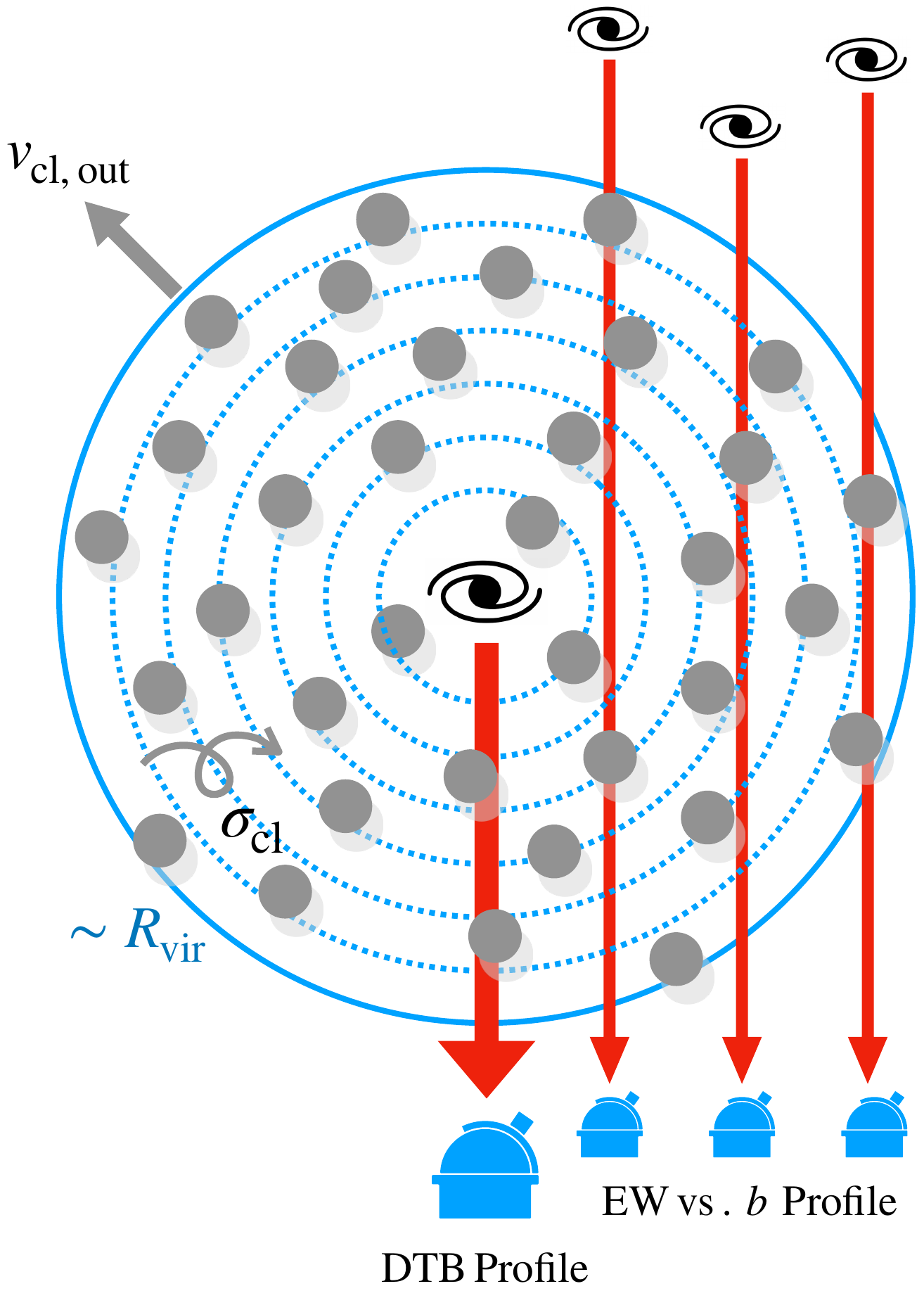}
    \caption{\textbf{Schematic for \textbf{ALPACA}, a non-Sobolev clumpy model for metal absorption lines.} For the DTB absorption line profile, it is assumed that a central source emits continuum photons isotropically and that all photons travel radially in a spherical halo that contains a number of absorbing clumps. For computational convenience, the halo is divided into a series of equally-spaced, concentric shells. The probability of escape for a continuum photon observed at a particular velocity is determined by the product of transmission probabilities through all radial shells. In each shell, the transmission probability is the sum of the probabilities of propagating through ``holes'' that are not occupied by any clumps (given by 1 $-\,C_{\rm f}(r_i)$) and penetrating through clumps (given by $C_{\rm f}(r_i) e^{-\tau_{\rm ion} (v - v_i)}$). The EW v.s. $b$  profile can be similarly calculated at different impact parameters. We refer the readers to Section \ref{sec:formalism} for a detailed derivation. 
    \label{fig:schematic}}
\end{figure}

As illustrated in Figure \ref{fig:schematic}, we consider the escape of photons from an idealized, spherical halo filled by an ensemble of spherical clumps that contain the corresponding metal ions (e.g. Si$^+$ or C$^+$) that produce the absorption. For the sake of computational convenience, we assume a spherical halo with inner and outer boundaries defined by the clump launch  radius $r_{\rm min}$ and the halo extent $r_{\rm h}$, respectively; we then divide the halo into a series of concentric shells, equally-spaced in radius. The absorption contributed by all concentric shells constitutes the total absorption of the model. The interval between the midplanes of two adjacent radial shells is $d = (r_{\rm h} - r_{\rm min}) / {N_{\rm shell}}$, where $N_{\rm shell}$ is the total number of shells. The optimal way of choosing $N_{\rm shell}$ will be discussed later in this paper. 

The \texttt{ALPACA} model accounts for the non-zero width of the absorption cross section in the velocity space and does not use the commonly adopted Sobolev approximation, which assumes that absorption occurs only when a photon appears exactly at the line center in the reference frame of the absorbing gas. Instead, in \texttt{ALPACA}, each outgoing photon will suffer from absorption by clumps with a range of velocities, even if it is not at the line center (i.e. out of resonance) in the reference frame of a clump. As we will demonstrate later in Section \ref{sec:non_sobolev}, this is particularly important when the velocity gradient of the clumps is small or the clump random motion is non-negligible. To escape, each photon must pass through every shell consecutively. In each shell, a photon may either pass freely through ``holes'' where no clump exists (with a probability of $1 - C_{\rm f}$, where $C_{\rm f}$ is the geometric covering fraction of the clumps), or penetrate though a clump (with a probability of $C_{\rm f}e^{-\tau_{\rm ion}}$, where $\tau_{\rm ion}$ is the optical depth of one clump of the relevant transition). Therefore, the probability of escape for a photon originating from the ISM of a galaxy can be expressed as:

\begin{equation}
P_{\rm esc}(-v) = \prod_{i=1}^{N_{\rm shell}} (1 - C_{\rm f}(r_i) + C_{\rm f}(r_i) e^{-\tau_{\rm ion} (v - v_i)})
\label{eq:pesc}
\end{equation}
Here $-v$ represents the location in the rest-frame velocity space. This implies, e.g., if the clumps are outflowing with $v > 0$, absorption on the blue side at $-v < 0$ will be observed. $v_i$ is the (average) clump velocity in the $i$-th shell, determined by the clump radial velocity profile:

\begin{equation}
v_{i} = v_{\rm cl}(r) |_{r = r_i}
\label{eq:vcl_r}
\end{equation}
where $r_i (i = 1, 2, ..., N_{\rm shell})$ are the radial locations of the midplanes of all the shells where absorption will be calculated. $C_{\rm f}(r_i)$ is the clump geometric covering fraction at $r_i$, and $\tau_{\rm ion}(v - v_i)$ is the clump optical depth of the relevant transition evaluated at $v - v_i$, which is the photon's apparent frequency in velocity space in the reference frame of the clumps outflowing at $v_i$. 

\begin{figure}
\centering
\includegraphics[width=0.47\textwidth]{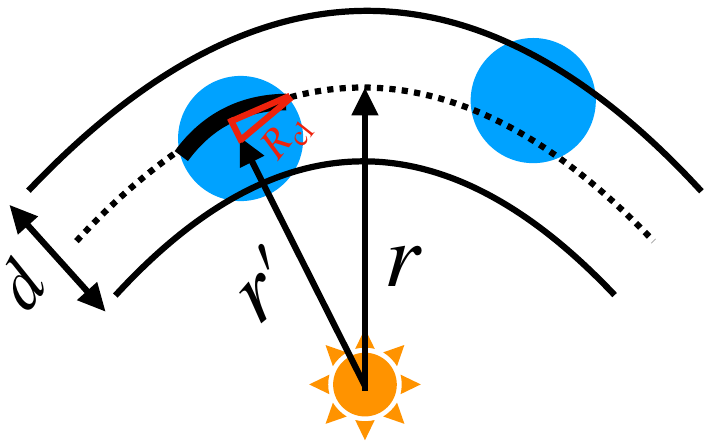}
    \caption{\textbf{Schematic for calculating the geometric covering fraction of clumps at radius $r$.} The contribution of one clump (as shown in blue) to the geometric covering fraction at a shell midplane at $r$ (as shown by the dotted black arc) is given by $\sim \pi [R^2_{\rm cl}(r')-(r-r')^2]$, which, after being integrated over $r \pm d/2$, gives the total geometric covering fraction at $r$, $C_{\rm f}(r)$.
    \label{fig:schematic_fc}}
\end{figure}

The geometric gas covering fraction $C_{\rm f}$, which is the fraction of the halo area covered by clumps at radius $r$, is given by (see also \citealt{Dijkstra12}):

\begin{equation}
C_{\rm f}(r)\approx \pi \int_{r-\frac{d}{2}}^{r+\frac{d}{2}}dr' n_{\rm cl}(r')[R^2_{\rm cl}(r')-(r-r')^2] \approx \pi n_{\rm cl}(r) [R^2_{\rm cl}(r)d - \frac{d^3}{12}]
\label{eq:fc}
\end{equation}
where $n_{\rm cl}(r)$ and $R_{\rm cl}(r)$ are the number density\footnote{Here the ``clump number density'' refers to the volumetric number density of the clumps, defined as the number of clumps per physical volume. Not to be confused with the number density of ions \emph{within} the clumps, which we do not use in the model (we consider the ion column density within the clumps instead).} and radius of the clumps at $r$, respectively. Eq. (\ref{eq:fc}) comes from the operation that each clump is assigned to a shell (with a thickness of $d$) within which its center is located and will only contribute to the absorption of this shell. Therefore, the clumps that contribute to the absorption of a particular shell must have their centers located within $r \pm d/2$. The way of calculating the contribution to $C_{\rm f}(r)$ for each clump is illustrated in Figure \ref{fig:schematic_fc}.

Recall that the clump optical depth can be written as:

\begin{equation}
\tau_{\rm ion} (v - v_i) = N_{\rm ion,\,cl}(r_i)\sigma_{\rm ion}(v - v_i)
\label{eq:tauv}
\end{equation}
where $N_{\rm ion,\,cl}(r_i)$ is the clump's ion column density\footnote{For a spherical clump, $N_{\rm ion,\,cl} = \frac{4}{3}n_{\rm ion,\,cl}R_{\rm cl}$, where $n_{\rm ion,\,cl}$ is the ion number density within the clump \citep{Gronke17b}.} at $r_i$, and $\sigma_{\rm ion}(v)$ is the cross section of the ion (both as a function of velocity), given by:

\begin{equation} 
\sigma_{\rm ion}(v) = \frac{\sqrt{\pi} e^2
  f_{\rm line}}{m_ec\Delta\nu_{\rm D}} H(a,x)
\label{eq:sigma}
\end{equation}
where $e$ is the electron charge, $m_e$ is the electron mass, $f_{\rm line}$ is the oscillator strength of the line transition, $\Delta \nu_{\rm D} = b_{\rm D}\nu_0/c$ is the Doppler width, $b_{\rm D}$ is the Doppler parameter within a single clump, $a = A_{\rm line} / (4\pi\Delta\nu_{\rm D})$ is the normalized natural line width (where $A_{\rm line}$ is the Einstein coefficient of the transition), $x = (\nu - \nu_{0})/\Delta\nu_{\rm D} = -v/b_{\rm D}$ is the unitless photon frequency, and $H(a,x)$ is the Voigt function:

\begin{equation}
H(a,x) = \frac{a}{\pi} \int_{-\infty}^{+\infty} \frac{e^{-y^2}}{(y-x)^2 + a^2}\mathrm{d}y
\label{eq:Voigt}
\end{equation}

Next, specific radial profiles can be assumed for clump outflow velocity, clump number density, clump ion column density and clump radius as a function of the clumps' galactocentric radius $r$, namely $v_{\rm cl}(r)$, $n_{\rm cl}(r)$, $N_{\rm ion,\,cl}(r)$ and $R_{\rm cl}(r)$, respectively. The values of $n_{\rm cl}(r)$ and $R_{\rm cl}(r)$ can be used to calculate $C_{\rm f}(r)$ using Eq. (\ref{eq:fc}), whereas $v_{\rm cl}(r)$ and $N_{\rm ion,\,cl}(r)$ can be used to calculate $\tau_{\rm ion} (v)$ using Eq. (\ref{eq:tauv}) -- (\ref{eq:Voigt}).

Finally, with Eq. (\ref{eq:pesc}) to (\ref{eq:Voigt}), one can derive a (normalized) model absorption line profile, whose intensity is proportional to the photons' escape probability:

\begin{equation}
\frac{I(v)}{I_{\rm cont}} = P_{\rm esc}(v)
\label{eq:profile}
\end{equation}
where $I_{\rm cont}$ is the intensity level of the continuum.

\subsubsection{Absorption at Different Impact Parameters}\label{sec:abs_at_b}

\texttt{ALPACA} can also model the absorption of photons that are emitted from a background source at a particular impact parameter (i.e. along transverse sightlines), which is suitable for studying quasar-quasar/quasar-galaxy/galaxy-galaxy pairs (see e.g. \citealt{QPQ1, QPQ2, QPQ3,  Steidel10, QPQ4, QPQ6}). For example, the equivalent width (EW) of the absorption as a function of impact parameter can be predicted in a manner similar to the derivation above (see also \citealt{Dijkstra12}, which focused on \lya\ absorption).

The EW at a particular impact parameter $b$ is given by:

\begin{equation}
{\rm EW}(b) = \int_b \mathrm{d}\lambda (1 - e^{-\tau(\lambda)}) = \frac{1}{\nu_0} \int_b \mathrm{d}v (1 - e^{-\tau(v)})
\label{eq:EW_b}
\end{equation}
where $\nu_0$ is the line center frequency of the transition, and the integral is performed over the observed velocities $v$ where absorption is seen at impact parameter $b$. The transmission at a particular velocity, $e^{-\tau(v)}$, comes from the contribution of individual clumps along the transverse sightline at $b$. It can be calculated by separating the sightline into a number of line segments (which is analogous to separating the spherical halo into different shells):

\begin{equation}
e^{-\tau(v)} = \prod_{i=1}^{N_{\rm seg}} (1 - C_{\rm f,\,\parallel}(r_i) + C_{\rm f,\,\parallel}(r_i) e^{-\tau_{\rm ion} (v - v_{i,\,\parallel})})
\label{eq:e_tau_b}
\end{equation}
where the product is evaluated over all $N_{\rm seg}$ line segments. The galactocentric radii of the centers of the line segments constitute an array of $r_i$ where the clump quantities are evaluated. Assuming that the angle between the vector $-\vec{r}_i$ and the line of sight towards the observer is $\theta$ ($\in[{\rm arcsin}(b/{r_{\rm h}}), \pi/2]$), the clump covering fraction and the clump radial velocity projected along the line of sight at $r_i$, $C_{\rm f,\,\parallel}(r_i)$ and $v_{i,\,\parallel}$, are given by:

\begin{equation}
C_{\rm f,\,\parallel}(r_i) = f_{\rm c}(r_i) \Delta l
\label{eq:Cf_b}
\end{equation}

\begin{equation}
v_{i,\,\parallel} = v_{\rm cl}(r_i) {\rm cos}\theta
\label{eq:v_b}
\end{equation}
where $f_{\rm c}(r_i) = \pi n_{\rm cl}(r_i) R_{\rm cl}^2(r_i)$ is the clump covering {\it{factor}} at $r_i$ (see Eq. \ref{eq:fa} below in Section \ref{sec:fc_Fv}), $\Delta l = 2\sqrt{r_{\rm h}^2 - b^2} / N_{\rm seg}$ is the length of each line segment, and $v_{\rm cl}(r_i)$ is the clump radial velocity at $r_i$. Combined with Eq. (\ref{eq:tauv}) -- (\ref{eq:Voigt}), the equations presented above can be used to derive the EW of a particular transition as a function of the impact parameter $b$.

\subsection{A Practical Implementation}\label{sec:implementation}

In practice, the solution to the \texttt{ALPACA} model can be further simplified by assuming specific functional forms (e.g. power-laws) for the radial profiles of clump parameters. In this section, we explore a practical implementation of the model, so that it can be conveniently applied to model observational data.

\subsubsection{Clump Outflow Kinematics}\label{sec:vcl}

The cool clumps in a galactic outflow can be accelerated via a number of different mechanisms, including radiation pressure (e.g. \citealt{Murray05, Thompson05, Martin05}), ram pressure (e.g. \citealt{Murray05, Fujita09, Martin09, Sharma12, Thompson16}) from the hot wind, and cosmic rays (e.g. \citealt{Socrates08, Everett08, Dorfi12, Recchia16, Zweibel17, Mao18, Jacob18, Chan19, Quataert22a, Quataert22b}). Since these mechanisms are often dependent on multiple physical parameters and the clumps are likely to be accelerated by several mechanisms at the same time, the actual scaling of the acceleration force with radius is uncertain and difficult to determine observationally. For simplicity, we explore an $r^{-\alpha}$ acceleration force, where the power-law index $\alpha$ describes how fast the acceleration force drops with the galactocentric radius. For example, $\alpha = 2$ is an approximate scaling expected for acceleration due to optically thin radiation pressure, ram pressure, or cosmic rays\footnote{Such a scaling is derived for the acceleration force per unit area; for cool clumps that are in pressure equilibrium with a hot wind, one might expect $\alpha = 4/3$ \citep{Steidel10}.} \citep{Murray05, Socrates08, Martin09, Chisholm16}. We stress that the formalism of the \texttt{ALPACA} model is general and applicable to other radial scalings of the acceleration force.

\begin{figure*}
\centering
\includegraphics[width=0.325\textwidth]{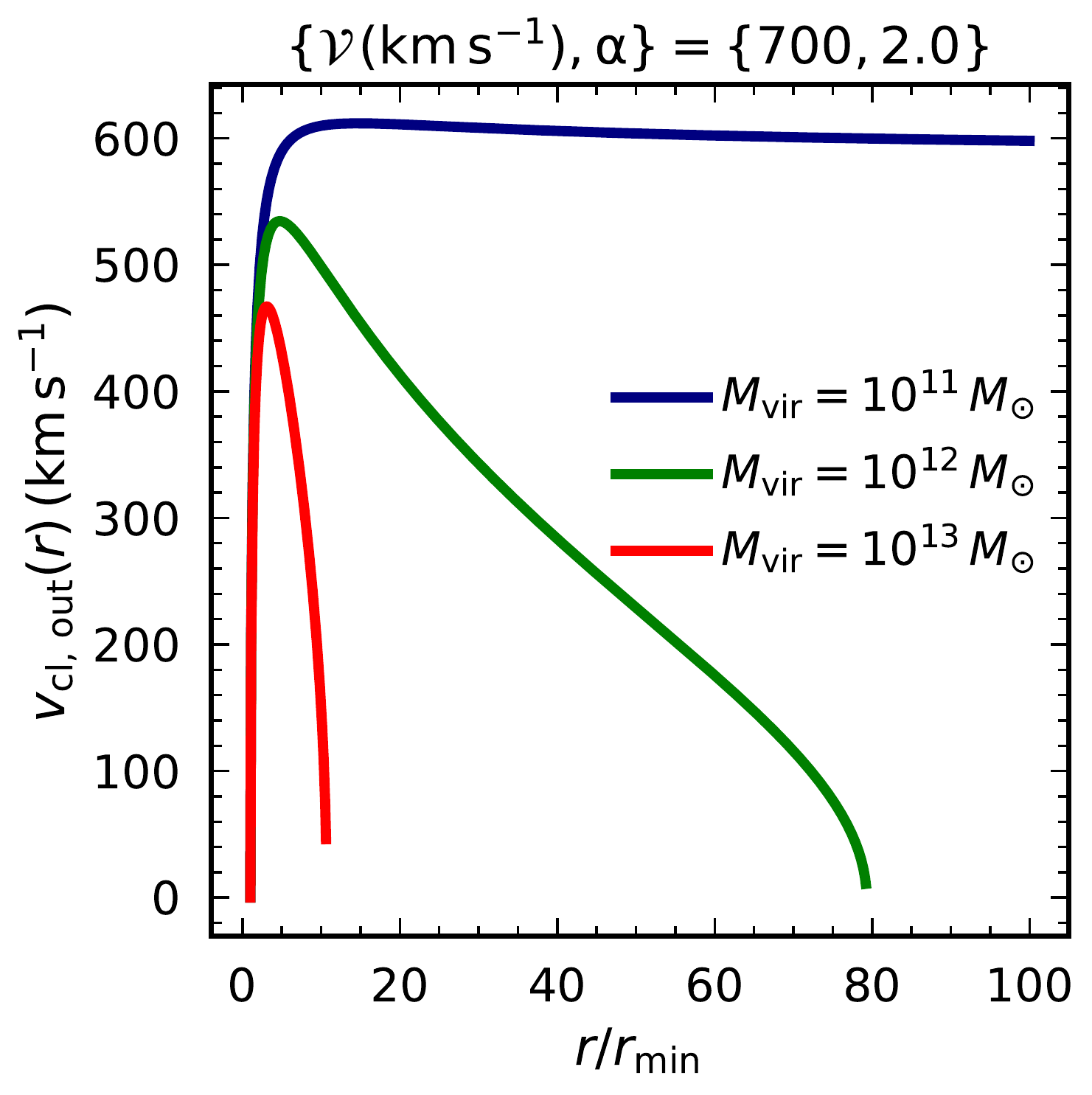}
\includegraphics[width=0.325\textwidth]{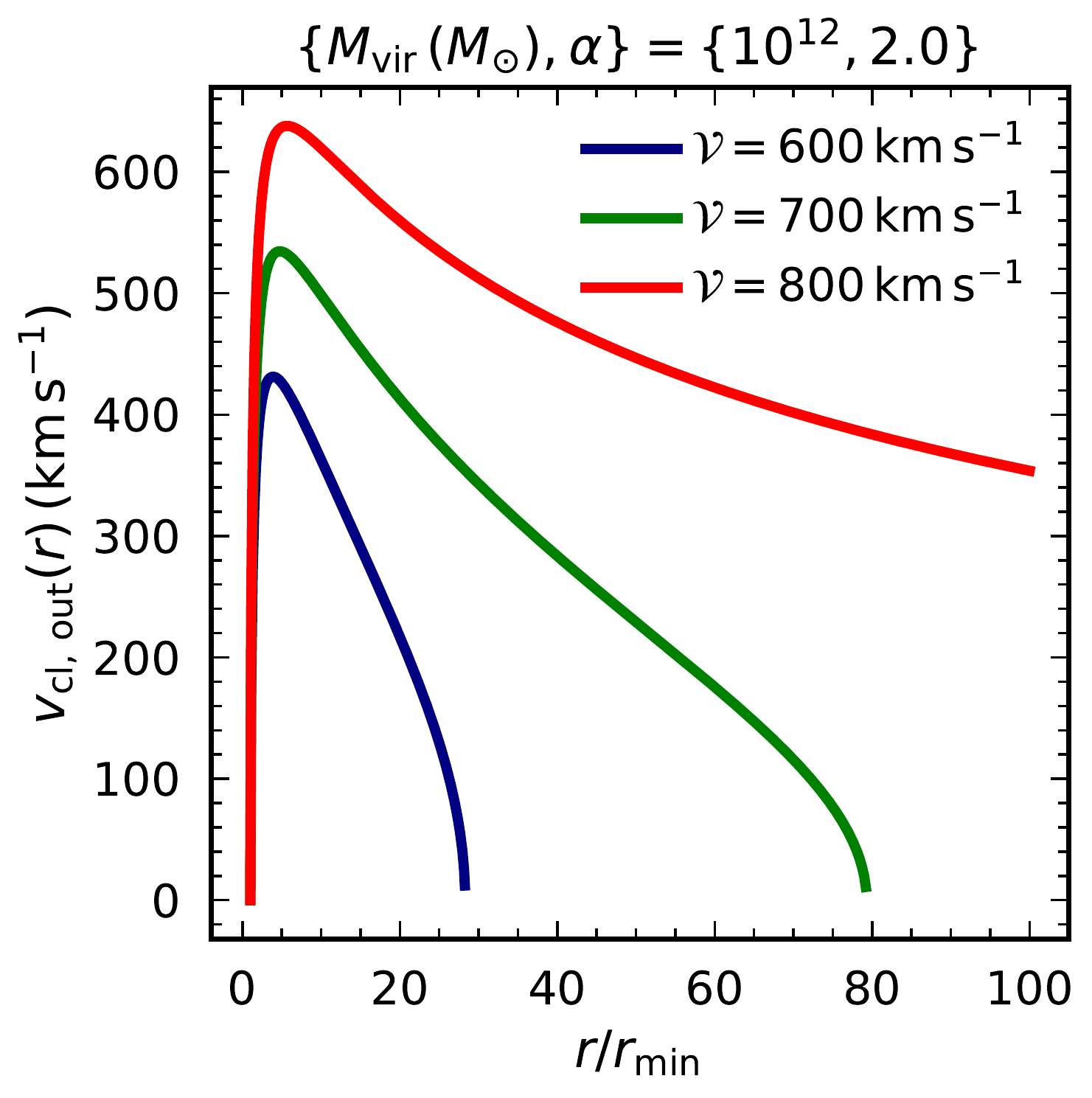}
\includegraphics[width=0.33\textwidth]{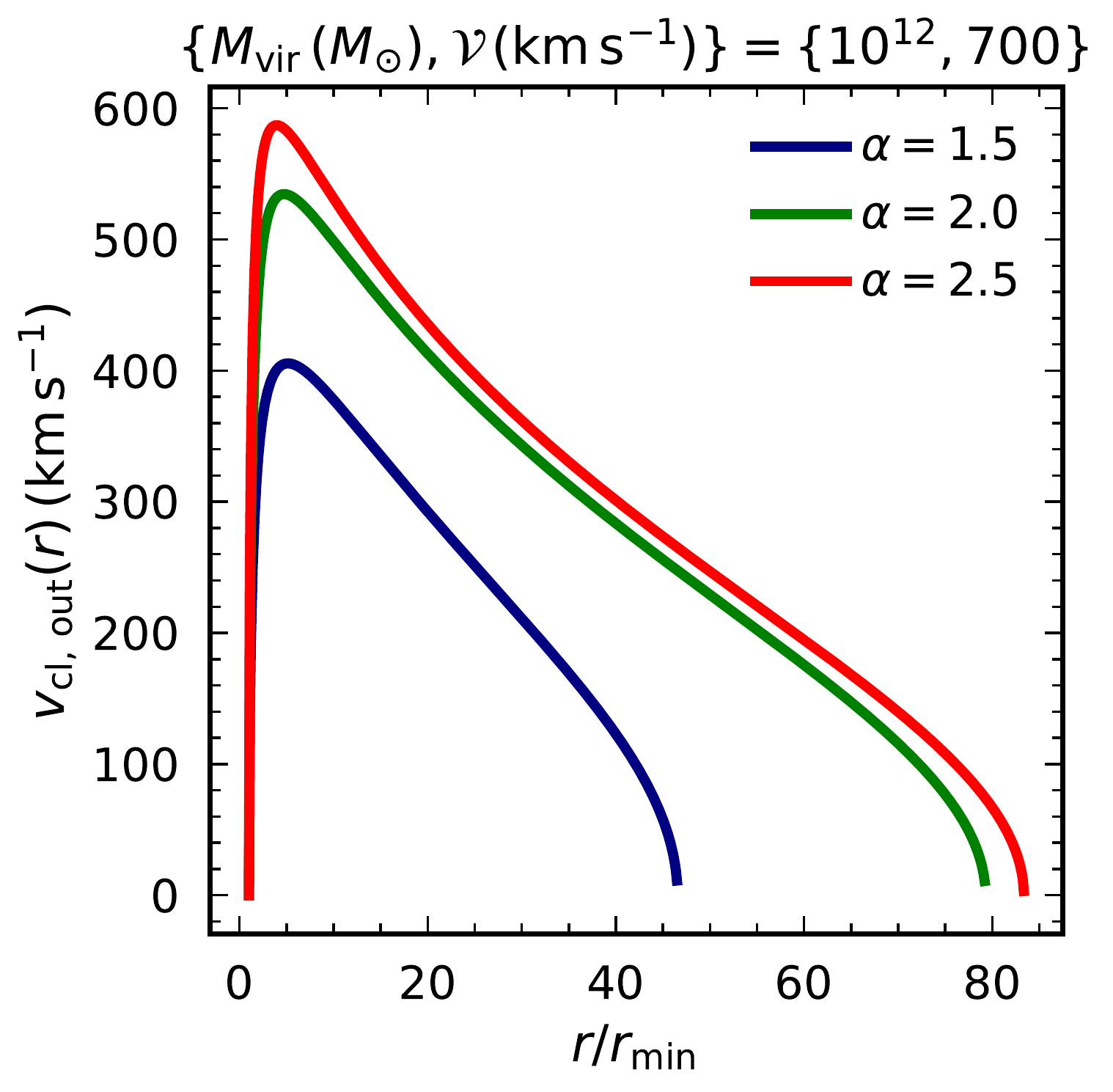}
    \caption{\textbf{Clump radial outflow velocity profiles $v_{\rm cl,\,out}(r)$ with different \{$M_{\rm vir}$, $\mathcal{V}$, $\alpha$\} values as given by Eq. (\ref{eq:v}).} In each panel, only one parameter is varied while the other two are fixed. The $v_{\rm cl,\,out}(r)$ profiles derived from varying one of the parameters are shown in different colors. 
    \label{fig:v_examples}}
\end{figure*}

In addition to acceleration, the outflowing clumps will inevitably suffer from gravitational deceleration from the mass of the dark matter halo. Therefore, the kinematic equation of an outflowing clump, is given by:

\begin{equation}
\frac{\mathrm{d}v_{\rm cl,\,out}(r)}{\mathrm{d}t}=-\frac{GM(r)}{r^2} + Ar^{-\alpha}
\label{eq:clump_momentum}
\end{equation}

Assuming a Navarro-Frenk-White (NFW, \citealt{Navarro95}) profile for the dark matter halo, the mass within radius $r$, is given by:

\begin{equation}
M(r) = 4\pi \rho_{0} r_{\rm s}^3 \displaystyle\left[\ln(1+r/r_{\rm s})-\frac{r/r_{\rm s}}{1+r/r_{\rm s}}\right]
\label{eq:Mr}
\end{equation}
where $\rho_{0}$ is the central density, given by:

\begin{equation}
\rho_0 = \frac{M_{\rm vir}}{4\pi r_{\rm s}^3\left[{\rm ln}(1+r_{\rm vir}/r_{\rm s})-\frac{r_{\rm vir}/r_{\rm s}}{1+r_{\rm vir}/r_{\rm s}}\right]}
\label{eq:rho0}
\end{equation}
where $M_{\rm vir}$ and $r_{\rm vir}$ are the halo virial mass and virial radius, respectively. $r_{\rm s} = r_{\rm vir} / c$ is the scale radius, where $c$ is the concentration parameter of the halo. In this context, Eq. (\ref{eq:clump_momentum}) can be further simplified as:

\begin{equation}
\frac{\mathrm{d}(\frac{1}{2}v^2_{\rm cl,\,out}(r))}{\mathrm{d}r}=-\frac{4\pi G\rho_{0} r_{\rm s}^3}{r^2}\left[{\rm ln}(1+r/r_{\rm s})-\frac{r/r_{\rm s}}{1+r/r_{\rm s}}\right] + \frac{A}{r_{\rm s}^{\alpha}}\Big{(}\frac{r_{\rm s}}{r}\Big{)}^{\alpha}
\label{eq:clump_momentum_2}
\end{equation}

Integrating the equation above from the clump launch radius $r_{\rm min}$ to $r$ yields the following solution:

\begin{equation}
\begin{aligned}
v_{\rm cl,\,out}(r) &= \Big{\{}\frac{2GM_{\rm vir}}{{\rm ln}(1 + c) - c/(1 + c)} \left[\frac{{\rm ln}(1+r/r_{\rm s})}{r}-\frac{{\rm ln}(1+r_{\rm min}/r_{\rm s})}{r_{\rm min}}\right] \\ 
& + \mathcal{V}^2 \left[1 - \Big{(}\frac{r}{r_{\rm min}}\Big{)}^{1-\alpha}\right]\Big{\}}^{1/2}
\end{aligned}
\label{eq:v}
\end{equation}
where we have replaced $A$ with $\mathcal{V}(\equiv\sqrt{2Ar_{\rm min}^{1-\alpha}/(\alpha-1)})$, which is the asymptotic outflow velocity if there were no gravitational deceleration.

Eq. (\ref{eq:v}) shows that $v_{\rm cl,\,out}(r)$ can be fully determined by six parameters in total: the virial mass $M_{\rm vir}$, the virial radius $r_{\rm vir}$, the concentration parameter $c$, the clump launch radius $r_{\rm min}$, the asymptotic velocity $\mathcal{V}$, and the power-law index $\alpha$. Among these parameters, $M_{\rm vir}$ and $r_{\rm vir}$ can be inferred via the stellar mass-halo mass relation (e.g. \citealt{Tasitsiomi04, Moster10, Moster13, Behroozi10, Behroozi13, Rodriguez15, Rodriguez17, Kravtsov18, Girelli20}) if the galaxy's stellar mass is known (e.g. from SED fitting), and $c$ can be inferred via the concentration-halo mass relation (e.g. \citealt{Wechsler02, Prada12, Dutton14, Ludlow14, Diemer15, Child18, Diemer19}). Therefore, for a given galaxy, this kinematic model has only three free parameters: $\mathcal{V}$, $r_{\rm min}$ and $\alpha$. In our following modeling, we simply fix $r_{\rm min}$ to 1 kpc as its effect on $v_{\rm cl,\,out}(r)$ is relatively minor. In Figure \ref{fig:v_examples}, we show several example $v_{\rm cl,\,out}(r)$ profiles by varying $M_{\rm vir}$, $\mathcal{V}$ and $\alpha$ individually (assuming $r_{\rm min} = 1$ kpc).

\subsubsection{Clump Number Density and Radius}\label{sec:ncl_rcl}

Heuristically, the clump number density $n_{\rm cl}(r)$ and the clump radius $R_{\rm cl}(r)$, can be assumed to vary radially in the form of a power-law:

\begin{equation}
n_{\rm cl}(r) = n_{\rm cl,\,0} \Big{(}\frac{r}{r_{\rm min}}\Big{)}^{-\gamma}
\label{eq:ncl}
\end{equation}

\begin{equation}
R_{\rm cl}(r) = R_{\rm cl,\,0} \Big{(}\frac{r}{r_{\rm min}}\Big{)}^{\delta}
\label{eq:rcl}
\end{equation}
where $n_{\rm cl,\,0} = n_{\rm cl}(r = r_{\rm min})$ and $R_{\rm cl,\,0} = R_{\rm cl}(r = r_{\rm min})$. Although it is reasonable to assume that $\gamma \geq 0$ and $\delta \geq 0$ due to the increase in the volume of the halo and the decrease in ambient pressure at large $r$, we allow $\gamma$ and $\delta$ to be negative as other physical mechanisms may be at play in clump destruction, fragmentation and (re)formation.

Next, we normalize $n_{\rm cl}$ properly by introducing the total volume factor of the halo, $F_{\rm V}$, which is the fraction of the halo volume occupied by the clumps. The total volume of the clumps in the halo, is given by:

\begin{equation}
\begin{aligned}
V_{\rm cl,\,total} & = \int_{r_{\rm min}}^{r_{\rm h}} 4 \pi r^2 n_{\rm cl}(r) V_{\rm cl}(r) \mathrm{d}r \\
& = \frac{16 \pi^2 n_{\rm cl,\,0} R_{\rm cl,\,0}^3 r_{\rm min}^3}{3 (3\delta+3-\gamma)} \Big{[}\Big{(}\frac{r_{\rm h}}{r_{\rm min}}\Big{)}^{3\delta+3-\gamma} - 1\Big{]}
\end{aligned}
\label{eq:Vcl_total}
\end{equation}

On the other hand, 
\begin{equation}
V_{\rm cl,\,total} = F_{\rm V} V_{\rm h} = F_{\rm V} \frac{4}{3} \pi (r_{\rm h}^3 - r_{\rm min}^3)
\label{eq:Vcl_total_fv}
\end{equation}

Eq. (\ref{eq:Vcl_total}) and (\ref{eq:Vcl_total_fv}) can be used to further simplify the expression for $C_{\rm f}(r)$ (Eq. \ref{eq:fc}) to:

\begin{equation}
\begin{aligned}
C_{\rm f}(r) & \approx \pi n_{\rm cl}(r) R^2_{\rm cl}(r) d\\
& = \frac{(3\delta+3-\gamma) F_{\rm V} \big{[}\big{(}\frac{r_{\rm h}}{r_{\rm min}}\big{)}^{3} - 1\big{]}}{4R_{\rm cl,\,0}\big{[}\big{(}\frac{r_{\rm h}}{r_{\rm min}}\big{)}^{3\delta+3-\gamma} - 1\big{]}}\Big{(}\frac{r}{r_{\rm min}}\Big{)}^{2\delta-\gamma}d\ 
\label{eq:cf_rcl}
\end{aligned}
\end{equation}

Eq. (\ref{eq:cf_rcl}) implies that there is a triple degeneracy among $F_{\rm V}$, $\delta$ and $\gamma$ -- specifically, a parameter set \{$F_{\rm V}$, $\gamma$, $\delta$\} gives an identical $C_{\rm f}(r)$ profile to the following parameter set (where $\Delta$ represents a particular variation in $\gamma$):

\begin{equation}
\{\frac{\big{[}\big{(}\frac{r_{\rm h}}{r_{\rm min}}\big{)}^{3\delta+3-\gamma+\Delta/2}-1\big{]}(3\delta+3-\gamma)}{\big{[}\big{(}\frac{r_{\rm h}}{r_{\rm min}}\big{)}^{3\delta+3-\gamma}-1\big{]}(3\delta+3-\gamma+\Delta/2)}F_{\rm V}, \gamma + \Delta, \delta + \frac{1}{2}\Delta\}
\label{eq:triple_degeneracy}
\end{equation}

As the model absorption lines are only sensitive to the $C_{\rm f}(r)$ profiles rather than the individual values of $F_{\rm V}$, $\gamma$ or $\delta$, in the following modeling, we simply fix $\delta = 0$ while keeping $F_{\rm V}$ and $\gamma$ as free parameters in order to reduce the parameter degeneracies and computational cost. The readers should keep in mind that in reality, it is likely that the clump radius varies with the galactocentric radius; nevertheless, such an effect is indistinguishable from a change in the radial distribution of the clumps under the current formalism of \texttt{ALPACA}.

\subsubsection{Number of Shells}\label{sec:Nshell}

Although in principle, the choice of the spacing between two adjacent shells $d$ is arbitrary, it is advantageous to choose a relatively small $d$ to better sample the radial velocity profile and improve the accuracy of the model. In Appendix \ref{sec:convergence}, we show that for $\sigma_{\rm cl} \lesssim 100 \kms$, the model converges at $d / R_{\rm cl} \sim 0.1$. Therefore, in the next section, we adopt $d / R_{\rm cl} \sim 0.1$ as it achieves sufficient accuracy with reasonable computational cost.

\section{Model Validation}\label{sec:valid}

\begin{figure*}
\centering
\includegraphics[width=\textwidth]
{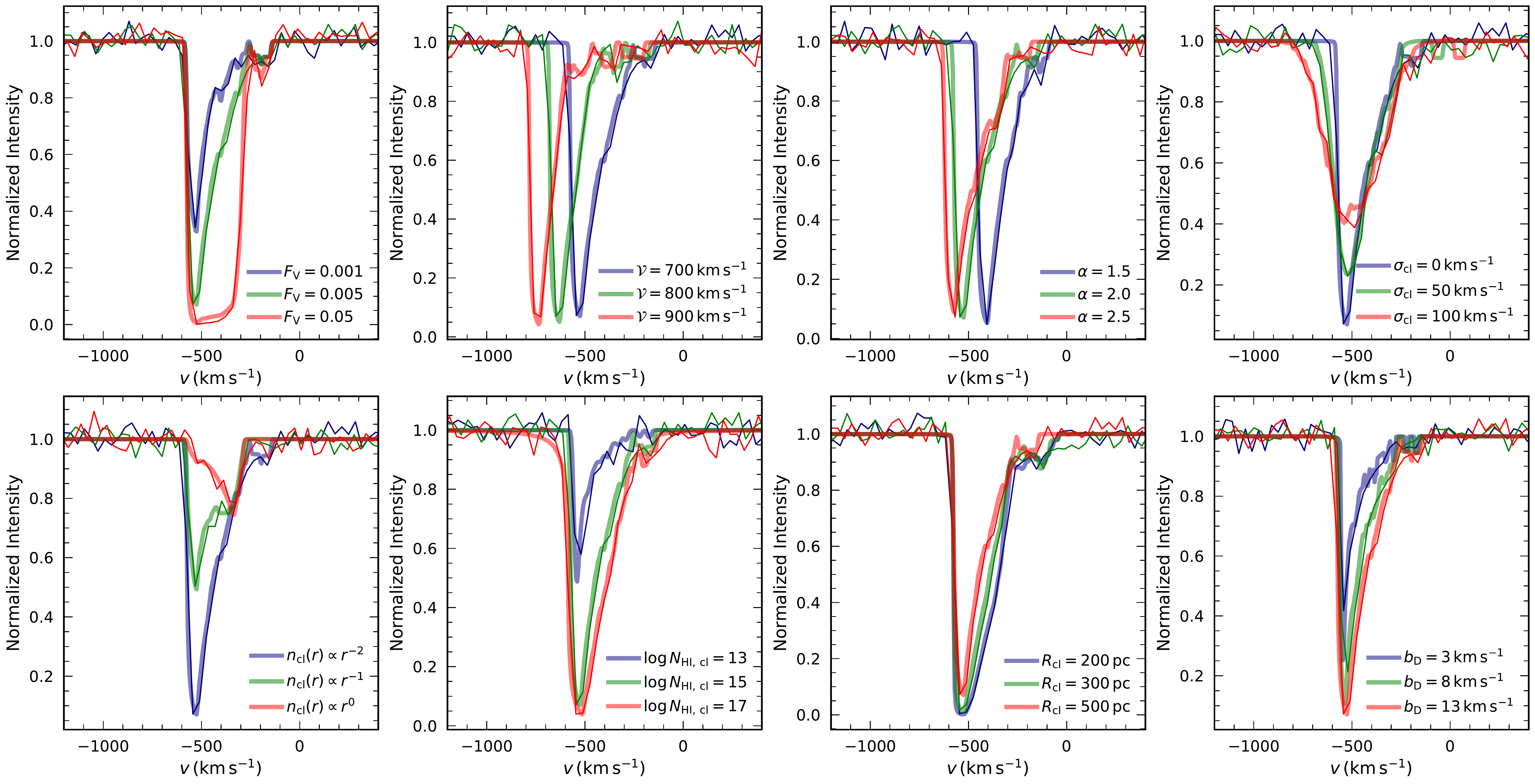}
    \caption{\textbf{Comparison of the absorption line profiles predicted by \texttt{ALPACA} and \texttt{tlac}.} The fiducial set of parameters are: $F_{\rm V} = 0.005, \mathcal{V} = 700 \kms, \alpha = 2.0, \sigma_{\rm cl} = 0 \kms, \gamma = 2.0,\,{\rm log}\,N_{\rm HI,\,cl} = 15,\,R_{\rm cl} = 500{\rm\,pc}, {\rm and}\,b_{\rm D} = 12.85\kms$. In each panel, only one parameter is varied (as indicated by three different colors) while the other parameters are fixed at their fiducial values. The model spectra predicted by \texttt{ALPACA} and \texttt{tlac} are shown in thick and thin curves, respectively. The absorption line profiles predicted by the two models are highly consistent over a wide range of physical parameters, suggesting that the formalism that we introduced in Section \ref{sec:formalism} is remarkably successful.}
    \label{fig:validation}
\end{figure*}

In this section, we test the validity of the \texttt{ALPACA} model with a Monte-Carlo radiative transfer (RT) code, \texttt{tlac} \citep{Gronke14,Gronke15}. \texttt{tlac} is specifically designed for simulating the RT process of \lya\ photons with idealized configurations. Nevertheless, the RT processes of \lya\ and metal lines (e.g. \SiII\ $\lambda$1260 and \CII\ $\lambda$1334) are very similar in nature, despite the following two subtle differences:

(1) The line cross section is different for different transitions. Such a difference can be easily accounted for by replacing the relevant coefficients and physical constants used in calculating the line cross section, namely the oscillator strength of the line transition $f_{\rm line}$, the Einstein coefficient of the transition $A_{\rm line}$, the wavelength of the transition $\lambda_{\rm line}$, and the ion mass $m_{\rm ion}$;

(2) Unlike \lya, the metal lines often have nearby non-resonant transitions (as the ground state is split into ${}^2P_{1/2}$ and ${}^2P_{3/2}$), so that a significant portion of the absorbed resonant photons can be re-emitted as non-resonant emission (e.g. ${\rm Si\,{\textsc {ii}}}^{\star}$ $\lambda$1265). We neglect such fluorescent emission for the moment as we are mostly focused on the absorption line profile in this work.

With those in mind, we can test whether the \texttt{ALPACA} model gives the correct absorption line profile with different clump radial velocity profiles, clump number density profiles, clump radii and clump ion column densities as inputs, by comparing with the Monte-Carlo RT simulations performed by \texttt{tlac} using the \lya\,$\lambda$1216 line. Once the model is validated, we can use it to predict other metal (e.g. Si and C) absorption line profiles by simply switching to a different transition.

In order to perform Monte-Carlo RT simulations, \texttt{tlac} requires several input parameters / radial profiles, namely the total clump volume filling factor $F_{\rm V}$, the radial distribution of the clumps, the clump radial velocity (including outflow and random motion), the clump column densities, and the clump radii. After the parameters of the clumps are fully specified, in each model, a UV continuum source is placed at the center of a spherically symmetric halo that emits 10$^5$ photons in the form of a flat continuum within $\pm 1500 \kms$ of the rest-frame wavelength of the \lya\ transition (1215.67\,\AA). All the photons will eventually escape from the halo (as we only consider a dust-free medium in this work), whereas a fraction of the emitted photons will be resonantly scattered by the clumps by one or more times before they escape.

For each Monte-Carlo RT simulation, all the photons that are scattered by the clumps at least once are filtered out and only the photons that have zero scatterings are used to construct the model absorption line profile\footnote{This is essentially assuming all the scattered photons will not re-enter the line of sight of the observer in a real observation.}. In this way, the output absorption line profile from \texttt{tlac}, which does not account for the contribution of re-emission from scattered photons (see Section \ref{sec:caveats} for a discussion of such re-emission and the associated ``infilling'' effect), can be directly compared to that of \texttt{ALPACA}.

Our tests are based on the practical implementation described in Section \ref{sec:implementation}, and are performed by varying the following key parameters or radial profiles, one at a time, with respect to the fiducial set of parameters ($F_{\rm V} = 0.005, \mathcal{V} = 700 \kms, \alpha = 2.0, \sigma_{\rm cl} = 0 \kms, \gamma = 2.0, {\rm log}\,N_{\rm HI,\,cl} = 15, R_{\rm cl} = 500\,{\rm pc}, b_{\rm D} = 12.85\kms$):
\begin{enumerate}
\item the total clump volume filling factor $F_{\rm V}$;

\item the clump radial velocity profile, including the clump outflow velocity $v_{\rm cl,\,out}(r)$ (which is a function of $\mathcal{V}$ and $\alpha$) and random velocity $v_{\rm cl,\,rand}(r)$. The total clump radial velocity $v_{\rm cl}(r)$, is given by:

\begin{equation}
v_{\rm cl}(r) = v_{\rm cl,\,out}(r) + v_{\rm cl,\,rand}(r)
\label{eq:vtot}
\end{equation}
where 
\begin{equation}
v_{\rm cl,\,rand}(r) \sim \mathcal{N}(v, \mu = 0, \sigma = \sigma_{\rm cl})
\label{eq:vrand}
\end{equation}
is a random velocity field in the form of a normalized Gaussian distribution that is characterized by $\sigma_{\rm cl}$, the 1D macroscopic velocity dispersion among the clumps;

\item the shape of the clump number density profile, namely the power-law index $\gamma$ in Eq. (\ref{eq:ncl});

\item the clump \HI\ column density $N_{\rm HI,\,cl}$;

\item the clump radius $R_{\rm cl}$;

\item the Doppler parameter within a single clump $b_{\rm D}$.

\end{enumerate}

These tests are designed to verify the consistency of the absorption line profiles predicted by \texttt{tlac} and \texttt{ALPACA} over a wide range of physical parameters. We note that at present, \texttt{tlac} only supports radially-varying $v_{\rm cl}(r)$ and $n_{\rm cl}(r)$, but not $N_{\rm HI,\,cl}(r)$ or $R_{\rm cl}(r)$, i.e. the clump column density and radius cannot yet be varied continuously as a function of radius. These tests are therefore our first attempt to validate the \texttt{ALPACA} model with the currently available capabilities of \texttt{tlac}. In Section \ref{sec:data} where we apply \texttt{ALPACA} to observational data, we fix the clump radius to be constant (see the justification for such a choice in Section \ref{sec:implementation} above) and restrict ourselves to using constant clump ion column densities. We therefore consider the tests described above sufficient for the validation and application of \texttt{ALPACA} in this work.

The results of these validation tests are presented in Figure \ref{fig:validation}. In each test, we consider a $z \sim 3$ galaxy with a halo mass of $M_{\rm vir} \sim 10^{11.8} \msun$ and assume that the clump launch radius $r_{\rm min}$ = 1 kpc, the halo radius $r_{\rm h}$ = 100 kpc. It can be seen that the \texttt{ALPACA} model is highly consistent with the \texttt{tlac} model over a wide range of physical parameters, suggesting that the simple formalism that we introduced in Section \ref{sec:formalism} is remarkably successful at describing the absorption of photons.

\section{Effect of Individual Parameters}\label{sec:params}

\begin{table*}
    \centering
    \caption{Definitions and priors of the free parameters of \texttt{ALPACA} used to perform joint fitting.}
    \label{tab:params}
    \setlength{\tabcolsep}{3pt}
    \begin{tabular}{ccc}
    \hline\hline
    Parameter & Definition & Prior Range \\ 
     (1)  & (2) & (3) \\
    \hline
    $A_{\rm ISM}$ & Amplitude of the ISM absorption component & [0, 1]\\
    $\sigma_{\rm ISM}$ (km\,s$^{-1}$) & Standard deviation of the ISM absorption component & [50, 200]\\
    ${\rm log}\,F_{\rm V}$ & Clump total volume filling factor & [-4.0, -0.5]\\
    $\mathcal{V}$ (km\,s$^{-1}$) & Clump asymptotic outflow velocity & [300, 2000] \\
    $\alpha$ & Power-law index in the clump acceleration force profile & [1.05, 2] \\
    $\gamma$ & Power-law index in the clump number density / covering fraction profile & [-5, 5] \\
    \hline\hline
    \end{tabular}
    \begin{tablenotes}
    \item \textbf{Notes.} The definitions and prior ranges of the free parameters of \texttt{ALPACA} used to fit the composite DTB \CII\,$\lambda$1334 spectrum and the EW v.s. $b$ profile jointly. The columns are: (1) parameter name; (2) parameter definition; (3) prior range of the parameter.
    \end{tablenotes}
\end{table*}

Figure \ref{fig:validation} also illustrates the effects of different physical parameters in the \texttt{ALPACA} model, and we summarize them as follows:

\begin{itemize}
    \item Clump volume filling factor $F_{\rm V}$: Increasing $F_{\rm V}$ will increase the depth of and broaden the width of the flux minimum (``trough'') while keeping the location of the trough and the velocity range of the absorption profile roughly constant. This is because the clump covering fraction $C_{\rm f}(r)$ has increased proportionally at each radius (cf. Eq. \ref{eq:cf_rcl}). 
    \item Clump asymptotic outflow velocity $\mathcal{V}$: Modifying $\mathcal{V}$ simply shifts the overall spectrum horizontally without changing the shape of the profile. Note that the location of the trough corresponds to the maximum of $|v_{\rm cl}(r)|$, which is always smaller than $\mathcal{V}$ due to gravitational deceleration (cf. Eq. \ref{eq:v}). 
    \item Power-law index in the clump acceleration force profile $\alpha$: Similar to $\mathcal{V}$, changing $\alpha$ also shifts the spectrum horizontally, although in a rather non-linear way. As $\alpha$ increases, the maximum clump outflow velocity increases (see Figure \ref{fig:v_examples}), and the location of the trough shifts bluewards correspondingly.
    \item Clump radial velocity dispersion $\sigma_{\rm cl}$: Increasing $\sigma_{\rm cl}$ tends to reduce the depth of the trough and broaden the ``wings'' of the absorption line profile. This can be understood as an effective broadening in the range of the clump velocities that produces the absorption (cf. Eq. \ref{eq:vtot}).
    \item Power-law index in the clump number density profile $\gamma$: Decreasing $\gamma$, which yields a flatter radial declining profile for the number density of the clumps, tends to decrease the depth of the trough and shift the location of the trough nearer to the line center. This is because decreasing $\gamma$ reduces $C_{\rm f}(r)$ (cf. Eq. \ref{eq:cf_rcl}) and effectively places more clumps in the outer region of the halo where the clumps have decelerated to lower velocities.
    \item Clump \HI\ column density $N_{\rm HI,\,cl}$: Increasing $N_{\rm HI,\,cl}$ deepens the trough and broadens the wings of the absorption by increasing the clump optical depth at all velocities (cf. Eq. \ref{eq:tauv}). 
    \item Clump radius $R_{\rm cl}$: Increasing $R_{\rm cl}$ tends to decrease the depth of the trough, as it also changes $n_{\rm cl}(r)$ at a fixed $F_{\rm V}$ and the net effect is to decrease $C_{\rm f}(r)$ and produce less absorption (cf. Eq. \ref{eq:cf_rcl}). 
    \item Clump Doppler parameter $b_{\rm D}$: Increasing $b_{\rm D}$ yields more absorption at different velocities without increasing the observed velocity range of absorption, because the clump velocity distribution remains unchanged, yet there is more non-resonant absorption at each observed velocity.
\end{itemize}

\begin{figure*}
\centering
\includegraphics[width=\textwidth]{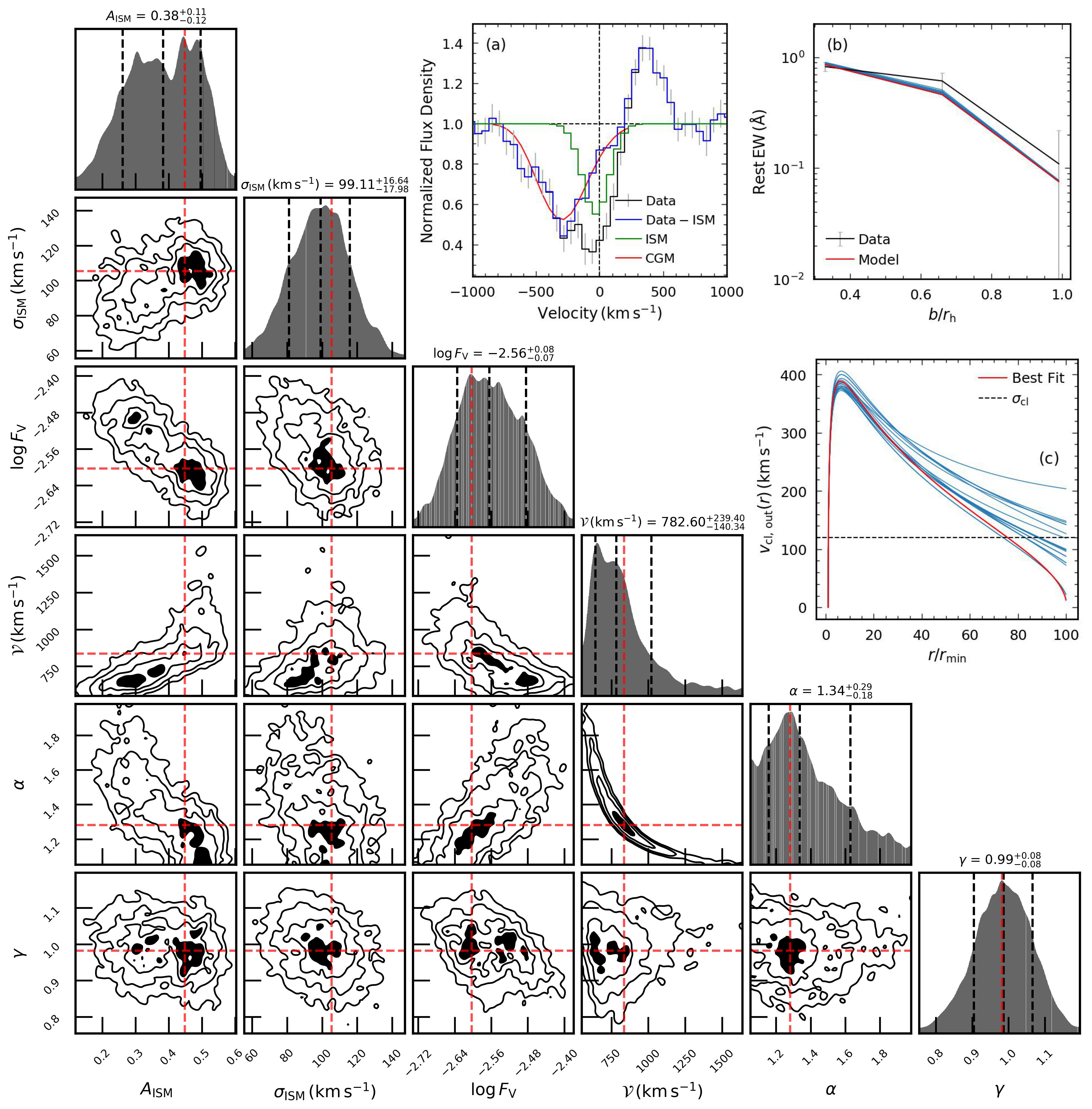}
    \caption{\textbf{Results of joint modeling the composite DTB absorption line profile and the EW v.s. $b$ profile of $\rm C\,{\textsc {ii}}\,\lambda$1334.} The posterior PDF is shown, along with the 1-$\sigma$ confidence intervals of the fitted parameters. The location of the maximum likelihood point is indicated by red dashed lines. On the upper right, panel (a) shows the best-fit model to the DTB absorption line profile. The non-outflowing ISM component and the outflowing CGM component are shown in green and red colors, respectively. Panel (b) shows the best-fit model (red) to the observed EW v.s. $b$ profile (black) at three different impact parameters: $b / r_{\rm h} \simeq [\frac{1}{3}, \frac{2}{3}, 1]$. Also shown are twenty models with the highest likelihoods (blue). Panel (c) shows the clump outflow velocity profiles of twenty models (blue) with the highest likelihoods in the parameter space, as well as the best-fit outflow velocity profile (red). The level of the clump radial velocity dispersion ($\sigma_{\rm cl} = 120 \kms$) is shown by a horizontal black dashed line. }
    \label{fig:best_fit}
\end{figure*}

\section{Application Example: Modeling the ISM \& CGM of a Sample of Lyman Break Galaxies}\label{sec:data}

Now that we have verified that the absorption line profiles predicted by the \texttt{ALPACA} model are reasonable by comparing them to the Monte-Carlo simulations carried out by \texttt{tlac}, next we apply \texttt{ALPACA} to the low-ionization, metal absorption lines observed in the rest-frame UV wavelengths. 

\subsection{Fitting the Composite DTB Spectrum and the EW v.s. $b$ Relation Simultaneously}\label{sec:fitting}

To tightly constrain the properties of the ISM and CGM of high-$z$ galaxies, we utilize both the DTB absorption line spectrum and the observed EW v.s. $b$ relation and model them simultaneously with \texttt{ALPACA}. 

\subsubsection{The Composite DTB Absorption Line Profile}\label{sec:dtb_data}

We use the stacked DTB \CII\,$\lambda$1334 spectrum of a sample of 55 (out of 124 in total) $z \sim$ 3 LBGs that are observed as part of the Keck Lyman Continuum Spectroscopic Survey (KLCS; \citealt{Steidel18}). The rest-UV spectra are obtained by the Low Resolution Imaging Spectrometer (LRIS) spectrograph on the Keck I telescope. This composite UV spectrum is constructed by stacking 55 individual spectra with accurate systemic redshift measurements (with uncertainties $<$ 20\,\kms) determined from nebular emission lines observed by MOSFIRE, which minimizes the effect of stochastic line-of-sight variation in the CGM and IGM attenuation compared to the spectrum of any single galaxy. The spectral resolution achieved at the wavelength of \CII\,$\lambda$1334 is $R \sim 1300$, or equivalently, FWHM $\sim 230\,\rm km\,s^{-1}$ or $\sigma \sim 98\,\rm km\,s^{-1}$. Before performing spectrum modeling, we have corrected the observed composite spectrum for CGM and IGM attenuation using the average transmission curve at $z \sim$ 3.05.

\subsubsection{The EW v.s. $b$ Profile}\label{sec:ew_b}

We use the observed EW v.s. $b$ relation of a sample of $z \sim$ 2 star-forming galaxy-galaxy pairs obtained by LRIS \citep{Steidel10}. The rest-frame EWs at three different impact parameters $\langle b \rangle$ = 31, 63, and 103 kpc are obtained by integrating over the corresponding stacked spectra of 42, 164, and 306 background galaxies, respectively. In addition, the EW at $b \sim 0$ is also estimated from the DTB spectra of all the foreground galaxies. We adopt the values given in Table 4 of \citet{Steidel10} for \CII\,$\lambda$1334 for our modeling.

\subsubsection{Joint Modeling of Two Datasets}\label{sec:joint_modeling}

To self-consistently model the DTB spectrum and the EW v.s. $b$ profile from two samples at different redshifts, we first check whether any correction needs to be applied to the datasets. We integrate the composite DTB \CII\,$\lambda$1334 absorption line profile of the $z \sim$ 3 LBG sample to derive a rest-frame EW (1.57 $\pm$ 0.03\,\AA) and compare with the average rest-frame EW at $b \sim 0$ measured from the foreground galaxies of the $z \sim$ 2 star-forming galaxy sample (1.72 $\pm$ 0.02\,\AA). We then apply a correction factor $f_{\rm corr} = 1.57 / 1.72 = 0.91$ to the three EWs measured at $b > 0$. In this way, we can model the two different datasets jointly as if they were both obtained at $z \sim$ 3. We note that the joint modeling we perform here is somewhat expedient; ideally one should do joint modeling on a sample with both $b = 0$ and $b > 0$ observations self-consistently.

In addition, we assume that there is a non-outflowing ISM component that also contributes to absorption on top of the clumpy, outflowing CGM component described above \citep{Steidel10}. The ISM absorption component is assumed to be a Gaussian centered at $v = 0$: $I_{\rm abs,\,ISM} = A_{\rm ISM}e^{-v^2/2\sigma_{\rm ISM}^2}$, where $A_{\rm ISM}$ and $\sigma_{\rm ISM}$ are the amplitude and standard deviation of the absorption, respectively. Note that $\sigma_{\rm ISM}$ and $\sigma_{\rm cl}$ are two independent parameters that characterize the gas velocity dispersion in the ISM and CGM, respectively.

To reduce the dimensionality of the parameter space, we take into account that the typical stellar mass of the $z \sim 3$ LBG sample is $M_{\star} \sim10^{9.7}$ \msun \citep{Pahl22}. Using the stellar mass-halo mass relation from \citet{Moster10} and the concentration-halo mass relation from \citet{Dutton14}, such a stellar mass corresponds to\footnote{We have adopted $H_0 = 70 \kms \rm{Mpc}^{-1}, \Omega_{\rm{m},\,0}=0.3$ and $\Omega_{\Lambda,\,0}=0.7$.} a virial mass of the halo $M_{\rm vir} \sim10^{12}$\,\msun, a virial radius $r_{\rm vir} \sim 76$ kpc, and a concentration parameter $c \sim 8.3$. For simplicity, in the model, we assume the halo radius $r_{\rm h}$ = 100\,kpc and the clump launch radius $r_{\rm min}$ = 1\,kpc. We remind the readers that the results are not sensitive to these choices. 

We further assume that the clump radius $R_{\rm cl}$ = 100\,pc \citep{Zahedy19}, clump C+ column density $N_{\rm C+,\,cl} = 10^{15}\,\rm cm^{-2}$ clump Doppler parameter $b_{\rm D} = 15 \kms$ (i.e. moderate internal turbulence), clump radial velocity dispersion $\sigma_{\rm cl} = 120 \kms$, which is close to the largest observed nebular emission line widths but slightly smaller than 1/$\sqrt{3}$ of the circular velocity of the halo that we consider. As a result, the \texttt{ALPACA} model used to jointly fit the composite DTB \CII\,$\lambda$1334 spectrum and the EW v.s. $b$ profile contains six parameters in total: the amplitude of the ISM absorption component $A_{\rm ISM}$, the standard deviation of the ISM absorption component $\sigma_{\rm ISM}$, the total clump volume filling factor $F_{\rm V}$ (Eq. \ref{eq:cf_rcl}), the clump asymptotic outflow velocity $\mathcal{V}$ (Eq. \ref{eq:v}), the power-law index in the clump acceleration force profile $\alpha$ (Eq. \ref{eq:v}), and the power-law index in the clump number density (or covering fraction) $\gamma$ (Eq. \ref{eq:cf_rcl}).

We use the nested sampling package \texttt{dynesty} \citep{Skilling04, Skilling06, Speagle20} in our fitting pipeline to map the posterior in such a multi-dimensional parameter space and find the best-fit parameters. At each sampled point in the parameter space, a model spectrum is calculated semi-analytically on-the-fly and convolved with the LRIS line spread function (LSF) with $\sigma \simeq 100\rm \,km\,s^{-1}$ before being compared to the input observed spectrum, and three EWs at $b = $ 33, 66, and 99 kpc are also calculated to be compared with the three observed EWs at $b > 0 $ correspondingly. The likelihood of each sampled point is the sum of the likelihoods of the model for the DTB spectrum and the EW v.s. $b$ profile. Each fitting run yields a posterior probability distribution function (PDF) of the six free model parameters. The uncertainties in the fitted parameters are determined as certain quantiles (e.g. 16\% -- 84\%, or 1-$\sigma$ confidence intervals) of the samples in the marginalized PDF. 

The priors of the parameters used for fitting are listed in Table \ref{tab:params}. In Figure \ref{fig:best_fit}, we present the best-fit parameters of the fitting run and the posterior PDF. We also present the best-fit DTB model absorption line profile, EW v.s. $b$ profile, and the clump radial outflow velocity profiles in three subpanels. 

\subsection{Interpreting the Modeling Results}\label{sec:best_fit_params}

We hereby examine the best-fit parameters of the model to understand the corresponding physical scenario. In the best-fit model, the ISM component is preferred to contribute significantly to the absorption near the line center, with a standard deviation of $\sigma_{\rm ISM} \sim 100 \kms$. Such a value is consistent with the nebular emission line widths convolved with the instrumental LSF. As for the CGM component, the clumps are preferred to be highly non-volume-filling ($F_{\rm V} \simeq 3 \times 10^{-3} \ll 1$), which corresponds to $f_{\rm c} = \frac{3}{2} F_{\rm V} \frac{r_{\rm h} - r_{\rm min}}{R_{\rm cl}} \simeq 4$ clumps with $R_{\rm cl} \sim 100\,\rm pc$ along each radial sightline. Such an $f_{\rm c}$ value implies that the halo is essentially fully covered by clumps to an external observer, as the likelihood for a radial sightline to contain zero clumps is $\sim e^{-4} < 2\%$. (see Eq. \ref{eq:fc_0_const} and \ref{eq:poisson} in Section \ref{sec:fc_Fv}). 

As shown in Figure \ref{fig:best_fit}, the clump radial velocities are preferred to be a superposition of outflow and velocity dispersion. The outflowing component has a rapid acceleration phase ($r / r_{\rm min} \lesssim 5$) towards a maximum outflow velocity of $v_{\rm out,\,max} \sim 400 \kms$ and then gradually decelerates until $r / r_{\rm min} \sim 100$. The location of the absorption trough basically corresponds to $-v_{\rm out,\,max}$, because the velocity gradient near $v = v_{\rm out,\,max}$ is close to zero and the number of clumps that provide resonant or nearly-resonant absorption at this velocity is the largest. The broad wings of the CGM absorption profile (especially on the blue side of the trough), however, are due to the perturbation on the clump outflow by a velocity dispersion of $\sigma_{\rm cl} = 120 \kms$. The total clump radial velocities range from $\sim -250 \kms$ to $\sim +700 \kms$, which is slightly narrower than the velocity range where significant absorption is seen ($v_{\rm obs} \sim -800 - 300 \kms$), because (1) the non-resonant absorption of clumps with $b_{\rm D} = 15\,\kms$ is accounted for; (2) the model spectrum is smoothed with $\sigma \simeq 100\,\kms$.

The best-fit power-law index in the clump acceleration force profile, $\alpha \simeq 1.3$, is consistent with the expected scaling ($\alpha = 4/3$) for cool clumps of constant mass that are in pressure equilibrium with a hot wind \citep{Steidel10}. The power-law index in the clump number density or covering fraction, $\gamma$, is preferred to be $\simeq 1$, which corresponds to a relatively steep decrease with radius. In general, at large $r$, the clump number density is expected to decrease due to the increase of the halo volume and the destruction of cold gas. On the other hand, the clumps are expected to expand in size due to the decrease in the pressure of the confining hot medium in the outer halo or grow due to various mixing and cooling processes. Our modeling suggests that the effect of the former physical process is more dominant over the latter.

Finally, we performed a fitting run by only fitting the DTB spectrum without using the EW measurements at $b > 0$. We find that in this case, $\gamma$ becomes poorly constrained, yet the values of the other five free parameters remain basically the same. Such an experiment emphasizes the importance of incorporating the information about the absorption at $b > 0$, which is to help constrain the radial profile of the clump number density and covering fraction.

\subsection{Parameter Degeneracies}\label{sec:param_degeneracy}

As is shown in the posterior distribution in Figure \ref{fig:best_fit}, there are a number of significant degeneracies between the parameters of the \texttt{ALPACA} model. Here we discuss them as follows:

\begin{itemize}
    \item $A_{\rm ISM}$ and $F_{\rm V}$: These two parameters are anti-correlated, as they contribute to the total absorption by modulating the amplitude of the ISM component and the clump covering fraction of the CGM component, respectively. 

    \item $A_{\rm ISM}$ and $\mathcal{V}$: These two parameters are positively correlated, as increasing $A_{\rm ISM}$ effectively adds more absorption around the line center and shifts the trough towards less negative velocities, whereas increasing $\mathcal{V}$ shifts the trough towards more negative velocities (see Figure \ref{fig:validation} in Section \ref{sec:params}).

    \item $\mathcal{V}$ and $\alpha$: These two parameters are anti-correlated, as a larger $\mathcal{V}$ increases the maximum clump outflow velocity, whereas a smaller $\alpha$ decreases the maximum clump outflow velocity (assuming $\mathcal{V}$ is fixed; see Figure \ref{fig:v_examples} in Section \ref{sec:vcl}). Such a degeneracy also translates to an anti-correlation between $A_{\rm ISM}$ and $\alpha$.

    \item $F_{\rm V}$ and $\gamma$: These two parameters are anti-correlated, as increasing $F_{\rm V}$ or decreasing $\gamma$ while keeping other parameters fixed results in an increase in the clump covering fraction and hence the total amount of absorption (see Eq. \ref{eq:cf_rcl} and Figure \ref{fig:validation}).

\end{itemize}

The parameter degeneracies of the \texttt{ALPACA} model may be broken with additional modeling or observations. For example, \lya\ emission modeling can be used to further constrain $\mathcal{V}$ and $\alpha$ and help break corresponding degeneracies, as the clump kinematic parameters are strongly correlated with particular \lya\ spectral features, e.g. the location of the double peaks and the blue-to-red peak ratio (\citealt{Li21b, Li21, Li21c, Erb22}).

\subsection{Model Anatomy: Where Does the Absorption Come from in the CGM?}\label{sec:model_anatomy}

\begin{figure}
\includegraphics[width=0.466\textwidth]{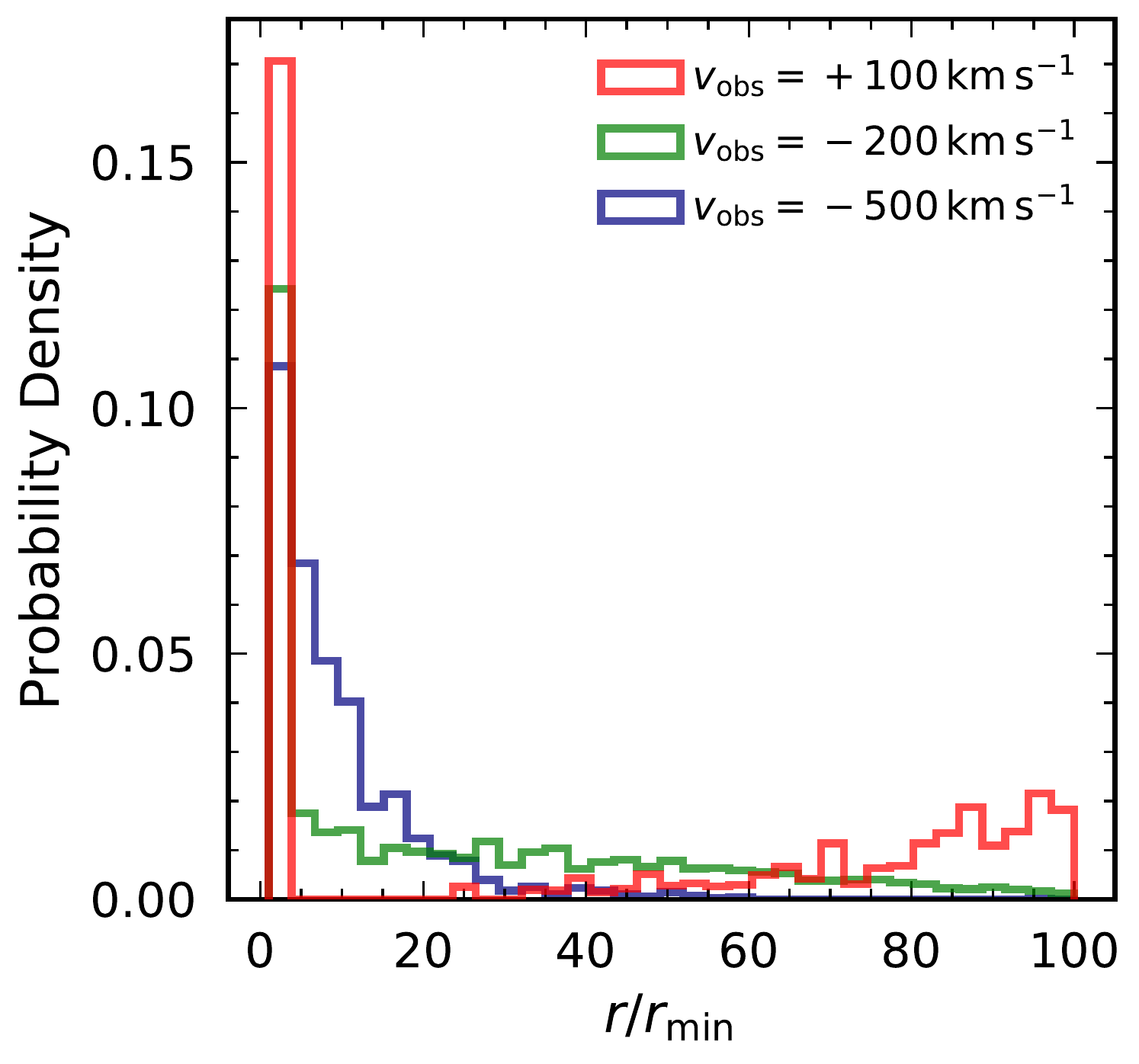}
    \caption{\textbf{Probability density distribution of the normalized galactocentric radii of the clumps, $r/r_{\rm min}$, weighted by the attenuation factor $C_{\rm f}(r) (1 - e^{-\tau(r)})$ at three different observed velocities.} For $v_{\rm obs} = -500\,\kms$, the probability density distribution decreases monotonically with radius, and the majority of absorption comes from $r / r_{\rm min} \lesssim 40$. For $v_{\rm obs} = -200\,\kms$, the contribution to the total absorption comes from all over the halo. For $v_{\rm obs} = +100 \kms$, the majority of absorption comes from $r / r_{\rm min} \sim 1$ and $r / r_{\rm min} \gtrsim 30$, whereas the contribution within $1 < r / r_{\rm min} < 30$ is negligible. }
    \label{fig:anatomy_plots}
\end{figure}

In \texttt{ALPACA}, since significant clump velocity dispersion is accounted for and there is no simple one-to-one mapping between the velocity space and the real space, it is not straightforward to describe where the majority of the absorption originates from. Therefore, here we zoom in on the internal structure of the model and reveal the relative contributions to the total absorption from the clumps located at different radii in the CGM. We examine three observed velocities in the DTB absorption line profile: $v_{\rm obs} = +100, -200, {\rm and}\, -500 \kms$.

As shown by Eq. (\ref{eq:pesc}), the ``attenuation'' factor of each shell, namely the fraction of flux density absorbed by the clumps, is given by $C_{\rm f}(r) (1 - e^{-\tau(r)})$, where $C_{\rm f}(r)$ and $\tau(r)$ are the clump covering fraction and optical depth at $r$, respectively. In Figure \ref{fig:anatomy_plots}, we plot the probability density distributions of the normalized galactocentric radii of the clumps, $r/r_{\rm min}$, weighted by the attenuation factor $C_{\rm f}(r) (1 - e^{-\tau(r)})$ at three observed velocities. In this way, we can clearly see where the clumps contribute most to the total absorption in the CGM.

In Figure \ref{fig:anatomy_plots}, we see that at all three different velocities, the largest contribution to the total absorption comes from $r / r_{\rm min} \sim 1$. This is because in the best-fit model, $C_{\rm f}(r) \propto n_{\rm cl}(r) \propto r^{-1}$, i.e. the clump number density or covering fraction peaks at $r / r_{\rm min} \sim 1$ and decreases fairly significantly with radius. For $v_{\rm obs} = -500\,\kms$, the probability density distribution decreases monotonically with radius, and the majority of absorption comes from $r / r_{\rm min} \lesssim 40$, within which the total velocity of the clumps is able to reach the corresponding resonant velocity $v_{\rm cl} = 500\,\kms$. For $v_{\rm obs} = -200\,\kms$, the contribution to the total absorption comes from all over the halo. For $v_{\rm obs} = +100 \kms$, the majority of absorption comes from $r / r_{\rm min} \sim 1$ and $r / r_{\rm min} \gtrsim 30$, whereas the contribution within $1 < r / r_{\rm min} < 30$ is negligible, although the clump number density is high. This is because the only location for the clumps to have a net negative total velocity $v_{\rm tot} = -100 \kms$ is where the clump kinematics are random velocity-dominated; i.e. $v_{\rm out} \simeq 0$ and $\sigma_{\rm cl} \simeq 120 \kms$, which is best satisfied at $r / r_{\rm min} \sim 1$ and $\sim 100$. 

The fact that the attenuation factor is highly velocity-dependent suggests that the clump optical depth $\tau(r)$ is still the dominant contributor to the total absorption, rather than the clump covering fraction $C_{\rm f}(r)$. Overall, compared to the models that do not account for significant clump random motion, \texttt{ALPACA} reveals a physical scenario where the absorption observed at a particular velocity is contributed by the clumps from a fairly broad range of radii, rather than from a single point of resonance. We will investigate these differences further in Section \ref{sec:non_sobolev}.

\subsection{Alternative Clump Radial Outflow Velocity Profiles}\label{sec:alternative_vout_r}

\begin{figure*}
\centering
\includegraphics[width=0.33\textwidth]
{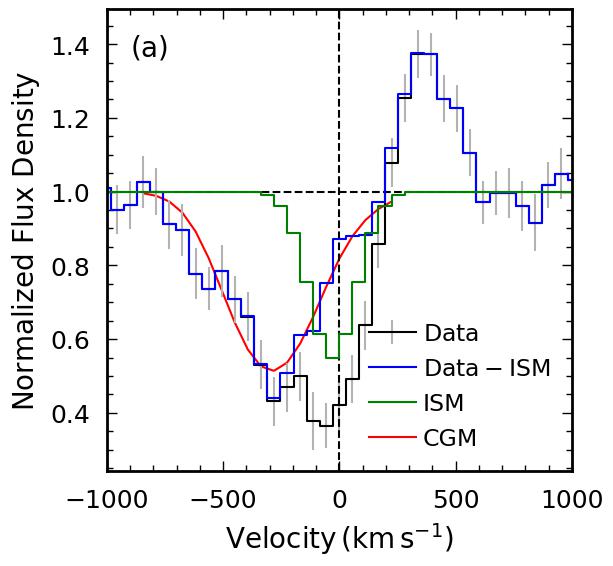}
\includegraphics[width=0.33\textwidth]
{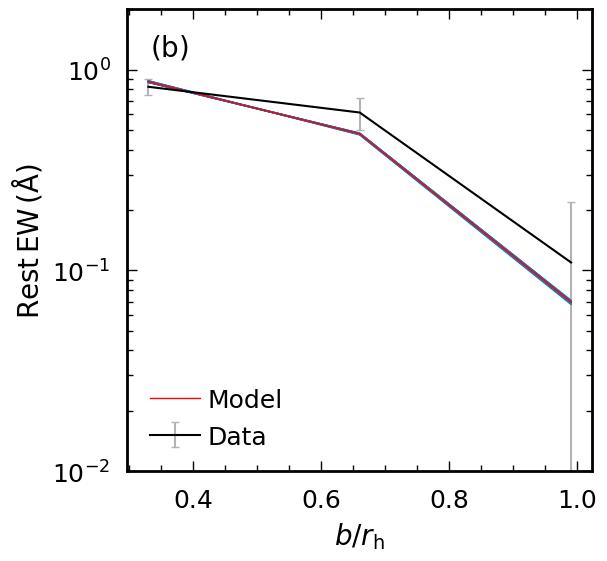}
\includegraphics[width=0.325\textwidth]
{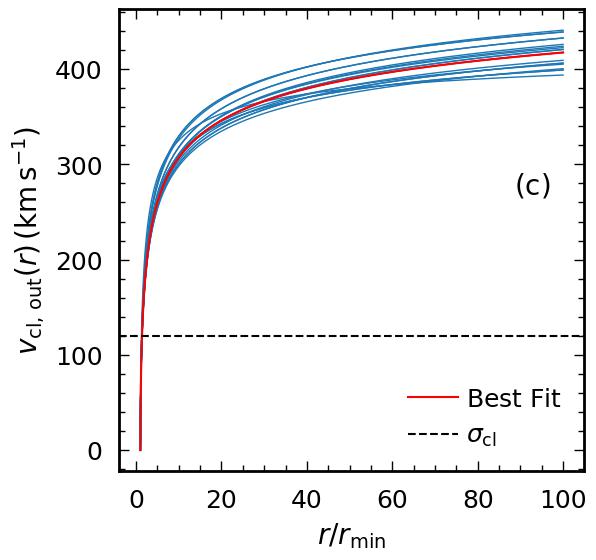}
    \caption{\textbf{Results of joint modeling using an alternative clump outflow velocity profile assuming gravitational deceleration is negligible (see Eq. \ref{eq:clump_momentum_S10}).} Panel (a) shows the best-fit model to the DTB absorption line profile. The non-outflowing ISM component and the outflowing CGM component are shown in green and red colors, respectively. Panel (b) shows the best-fit model (red) to the observed EW v.s. $b$ profile (black) at three different impact parameters: $b / r_{\rm h} \simeq [\frac{1}{3}, \frac{2}{3}, 1]$. Also shown are twenty models with the highest likelihoods (blue). Panel (c) shows the clump outflow velocity profiles of twenty models (blue) with the highest likelihoods in the parameter space, as well as the best-fit outflow velocity profile (red). The level of the clump radial velocity dispersion ($\sigma_{\rm cl} = 120 \kms$) is shown by a horizontal black dashed line. }
    \label{fig:best_fit_S10}
\end{figure*}

Although the formalism of the \texttt{ALPACA} model is general, any practical implementation for the purpose of application of the model will inevitably restrict the model to particular physical regimes. For example, the kinematic model of the clump outflow that we explored in Section \ref{sec:vcl} (Eq. \ref{eq:clump_momentum}) is highly simplistic and model-dependent, and will not capture all possible radial profiles of clump outflows. Therefore, here we explore a different type of radial profile for clump outflow velocities and see whether it can also provide a reasonable fit to the observational data.

We consider a scenario where the gravitational deceleration force is weak and negligible compared to the power-law acceleration force. In this case, the kinematic equation of an outflowing clump, is simply given by:

\begin{equation}
\frac{\mathrm{d}v_{\rm cl,\,out}(r)}{\mathrm{d}t} = Ar^{-\alpha}
\label{eq:clump_momentum_S10}
\end{equation}
which can be solved as:

\begin{equation}
v_{\rm cl,\,out}(r) = \mathcal{V} \Big{(}1 - \Big{(}\frac{r}{r_{\rm min}}\Big{)}^{1-\alpha}\Big{)}^{1/2}
\label{eq:v_S10}
\end{equation}
where we have replaced $A$ with $\mathcal{V}(\equiv\sqrt{2Ar_{\rm min}^{1-\alpha}/(\alpha-1)})$, the asymptotic clump outflow velocity at $r \rightarrow +\infty$. Eq. (\ref{eq:v_S10}) is exactly the radial velocity profile used by \citet{Steidel10}. We find that such a monotonically increasing radial outflow velocity profile is also able to yield a reasonable fit to the composite \CII\,$\lambda$1334 DTB spectrum and the EW v.s. $b$ profile modeled in Section \ref{sec:fitting} with the following best-fit parameters: $A_{\rm ISM} = 0.46^{+0.05}_{-0.06}, \sigma_{\rm ISM} = 103^{+14}_{-12} \kms, {\rm log}\,F_{\rm V} = -2.66^{+0.04}_{-0.04}, \mathcal{V} = 452^{+239}_{-90} \kms, \alpha = 1.30^{+0.40}_{-0.20}, \gamma = 1.05^{+0.08}_{-0.09}, {\rm log}\,N_{\rm C+,\,cl} = 15, \sigma_{\rm cl} = 120 \kms$. In Figure \ref{fig:best_fit_S10}, we show the best-fit models and the $v_{\rm cl,\,out}(r)$ profiles of this joint fitting run.

Such an experiment reminds us that there is still some freedom in the clump radial velocity distribution that may not be fully constrained by the joint fitting of the DTB spectrum and the EW v.s. $b$ profile. Nonetheless, these different radial velocity distributions do share one thing in common: they can all be decomposed into two velocity components -- a velocity dispersion and a radially-varying outflow, the latter of which is smaller by several hundred \kms\ than the maximum velocity of absorption. One promising way to further constrain the clump radial velocity profile is to incorporate spatially-resolved \lya\ emission modeling, assuming that the gas that produces LIS absorption lines is also responsible for producing extended \lya\ emission via resonant scattering. As the \lya\ blue-to-red peak flux ratio is sensitive to the local clump outflow velocity, one can distinguish whether the clump outflow has decelerated significantly or remains at a high speed at large radii by modeling the \lya\ profiles observed at the halo outskirts \citep{Erb22}. Recently work on mapping the 2D line-of-sight kinematics via \lya\ absorption may also help break the degeneracy \citep{Chen20}.

\section{Covering and Volume Filling Parameters of the Cool Gas}\label{sec:fc_Fv}

The physical properties of the ``cool'' ($T \sim 10^4 \rm K$) gas in a galactic environment, which is responsible for producing the UV absorption lines of the low ions, have been studied extensively in recent years, both theoretically and observationally. Notably, \citet{McCourt18} first carried out a comprehensive analysis by combining hydrodynamic simulations with observations and summarized with the following physical picture for the cool gas: a mist or fog of cloudlets with a large area covering \emph{factor}\footnote{Note that we have used a different terminology from \citet{McCourt18}. In this paper, we use the area covering \emph{factor} to refer to the average number of cloudlets intercepted per line of sight, and the covering \emph{fraction} to refer to the fraction of area covered by the clumpy gas. The volume filling fraction defined in \citet{McCourt18} has the same meaning as the volume filling factor defined in this work.} $f_{\rm c} \gg 1$ but a small total volume filling factor $F_{\rm V} \ll 1$ (see also \citealt{Liang20}). This physical picture is supported by a number of observational studies, e.g. \citet{Stocke13} report the volume filling factor of the cool clouds in the CGM of a sample of low-$z$ galaxies is on average a few percent (see also \citealt{Keeney17}), and \citet{Zahedy19} find that the mean volume filling factor of the cool gas is about $10^{-3}$ for massive ellipticals at $z \sim$ 0.4\footnote{See also \citet{Prochaska19}, who use FRB constraints and derive that the volume filling factor of the clumpy cool gas is $< 10^{-4}$ for a massive galaxy at $z \sim 0.4$.}. On the other hand, a close-to-unity coverage by the cool gas has been observed for the CGM halos of both galaxies and luminous quasars (e.g. \citealt{Prochaska13, Cantalupo14, Hennawi15, Borisova16, Cai17, Zahedy19, Rudie19}).

Before we move on, it is instructive to clarify the definition of the covering fraction $C_{\rm f}$, the covering factor $f_{\rm c}$, and the volume filling factor $F_{\rm V}$ of the cool gas. All these three parameters can be evaluated as either a global quantity of a halo or a radially-varying profile as a function of radius or velocity.

The expression for the covering fraction $C_{\rm f}$ as a function of radius has been derived in Eq. (\ref{eq:fc}). As for $f_{\rm c}$ and $F_{\rm V}$, considering the case of an idealized clumpy medium that only consists of spherical clumps and derived a relation between the covering factor and the volume filling factor. Specifically, the covering factor \emph{at a particular radius}, $f_{\rm c}(r)$, is given by \citep{Dijkstra12}:

\begin{equation}
f_{\rm c}(r) = n_{\rm cl}(r) \sigma_{\rm cl}(r) = n_{\rm cl}(r) \pi R_{\rm cl}^2(r)
\label{eq:fa}
\end{equation}
where $r$ is the radial location of the clumps, $n_{\rm cl}(r)$ is the number density of the clumps and $\sigma_{\rm cl}(r) = \pi R_{\rm cl}^2(r)$ is the geometric cross-section of a clump of radius $R_{\rm cl}(r)$. $f_{\rm c}(r)$ has the units of length$^{-1}$ and is analogous to the opacity $\kappa(r)$ in a homogeneous medium.

The volume filling factor \emph{at a particular radius}, $F_{\rm V}(r)$, is given by:

\begin{equation}
F_{\rm V}(r) = n_{\rm cl}(r) V_{\rm cl}(r) = n_{\rm cl}(r) \frac{4}{3} \pi R_{\rm cl}^3(r)
\label{eq:fv}
\end{equation}
where $V_{\rm cl}(r) = \frac{4}{3} \pi R_{\rm cl}^3(r)$ is the geometric volume of a clump with radius $R_{\rm cl}(r)$. Comparing with Eq. (\ref{eq:fc}) and (\ref{eq:fa}), we have:

\begin{equation}
F_{\rm V}(r) = \frac{4}{3} R_{\rm c}(r) f_{\rm c}(r)
\label{eq:three_relation}
\end{equation}

Note that although the above relation is derived under the assumption of spherical clumps, it also holds (modulo a geometric correction factor) for a more general geometric configuration of the clumpy gas. This is because $F_{\rm V}(r)$ and $f_{\rm c}(r)$ will always be proportional and different by a factor of $V_{\rm cl}(r) / \sigma_{\rm cl}(r)$, the clump volume-to-cross-section ratio.

One can further consider the following corresponding spatially-integrated quantities:

\begin{itemize}
    \item The \emph{total} volume filling factor of the halo, $F_{\rm V}$, is given by:

    \begin{equation}
    F_{\rm V} = \frac{1}{V_{\rm h}} \int_{r_{\rm min}}^{r_{\rm h}} F_{\rm V}(r) \mathrm{d}V(r)
    \label{eq:Fv}
    \end{equation}

    \item The \emph{integrated} gas covering factor, $f_{\rm c}$, i.e. the mean number of clumps along a line of sight at impact parameter $b$, is given by:

    \begin{equation}
    f_{\rm c}(b) = 2 \int_{b}^{r_{\rm h}} \frac{r \mathrm{d}r}{\sqrt{r^2-b^2}}f_{\rm c}(r) 
    \label{eq:fc_integrated}
    \end{equation}

    In particular, at $b = 0$ (``down the barrel''), $f_{\rm c}$ is given by:

    \begin{equation}
    f_{\rm c}(0) = 2 \int_{r_{\rm min}}^{r_{\rm h}}  f_{\rm c}(r) \mathrm{d}r = \frac{3}{2} \int_{r_{\rm min}}^{r_{\rm h}} \frac{F_{\rm V}(r)}{R_{\rm cl}(r)} \mathrm{d}r
    \label{eq:fc_0}
    \end{equation}
    which, in the special case where both $F_{\rm V}(r)$ and $R_{\rm cl}(r)$ are constant, can be simplified to:

    \begin{equation}
    f_{\rm c}(0) = \frac{3}{2} F_{\rm V} \frac{r_{\rm h} - r_{\rm min}}{R_{\rm cl}} 
    \label{eq:fc_0_const}
    \end{equation}

    It can be seen that in the limit of $r_{\rm h} / R_{\rm cl} \gg 1$, even a small $F_{\rm V} \ll 1$ can yield a large $f_{\rm c}(0) \gg 1$ \citep{McCourt18}.

    \item The \emph{integrated} gas covering fraction of the halo, $C_{\rm f}$, has the physical meaning of the fraction of sightlines at a particular impact parameter intercepted by at least one clump to an external observer. The Poisson probability of having sightlines at impact parameter $b$ that contain zero clumps is:

    \begin{equation}
    P(N_{\rm clump} = 0|f_{\rm c}(b)) = e^{-f_{\rm c}(b)}
    \label{eq:poisson}
    \end{equation}
    hence $C_{\rm f}(b) = 1 - P(N_{\rm clump} = 0|f_{\rm c}(b)) = 1 - e^{-f_{\rm c}(b)}$.

\end{itemize}

\begin{figure*}
\centering
\includegraphics[width=0.322\textwidth]{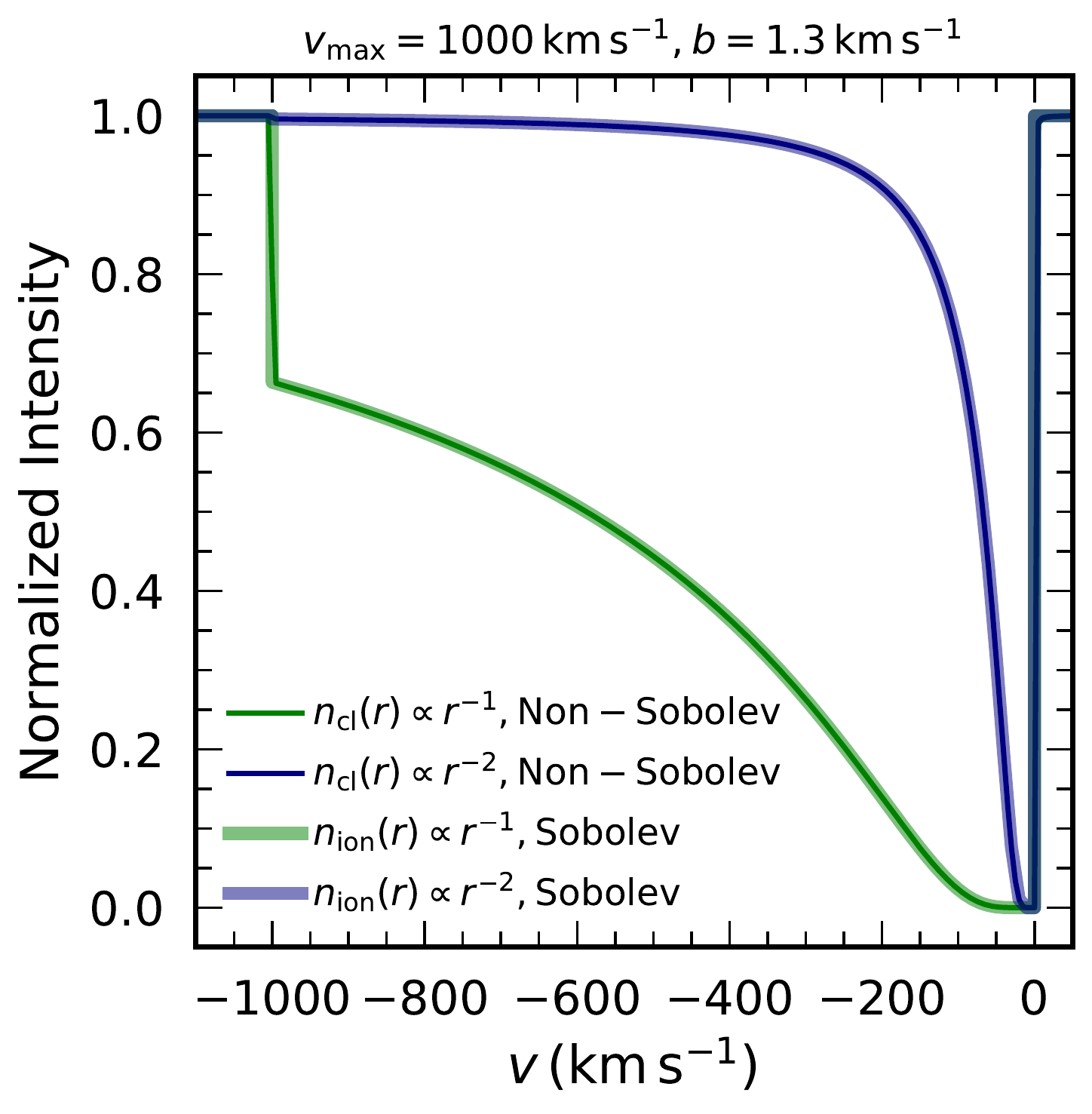}
\includegraphics[width=0.33\textwidth]{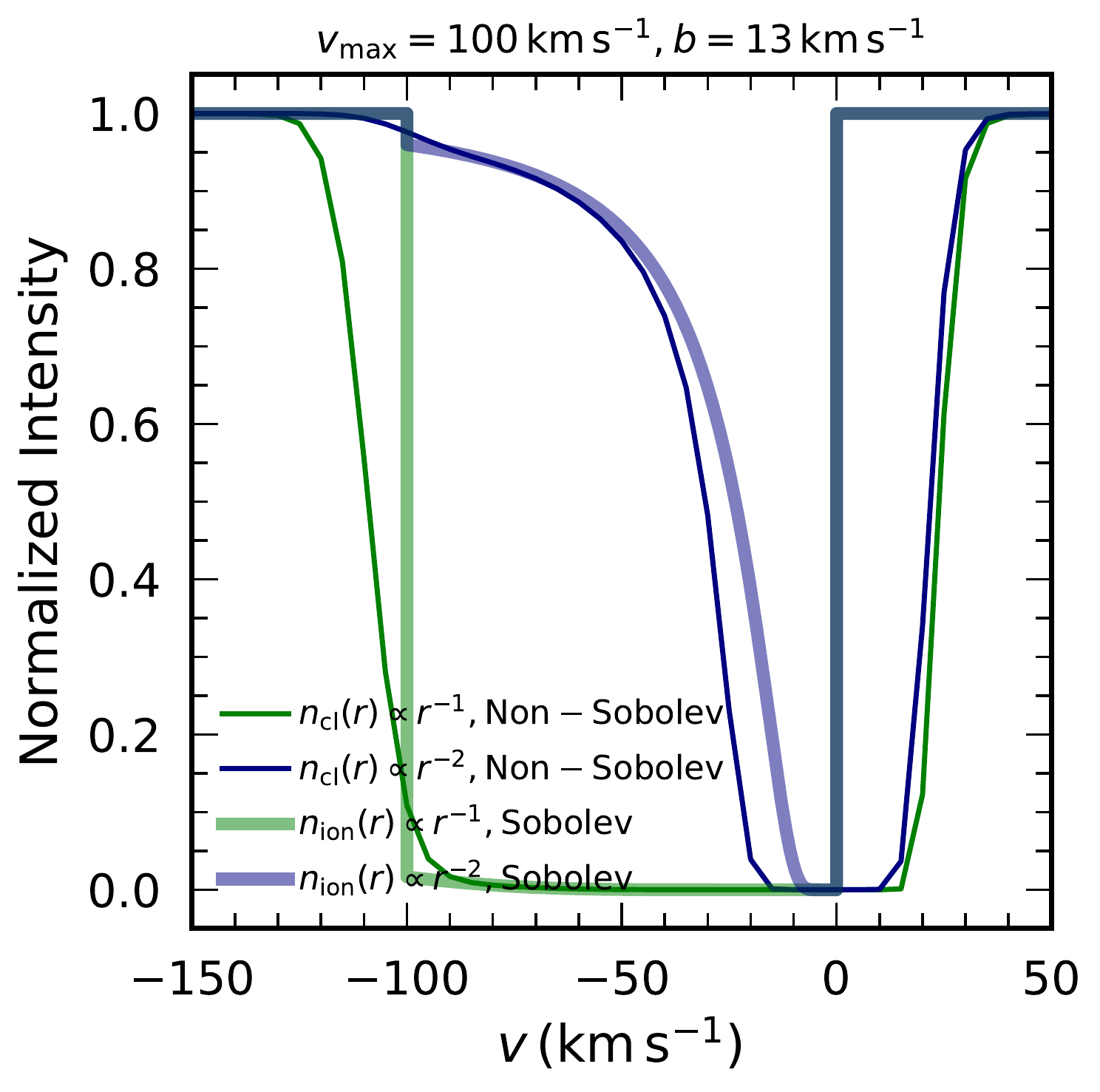}
\includegraphics[width=0.33\textwidth]{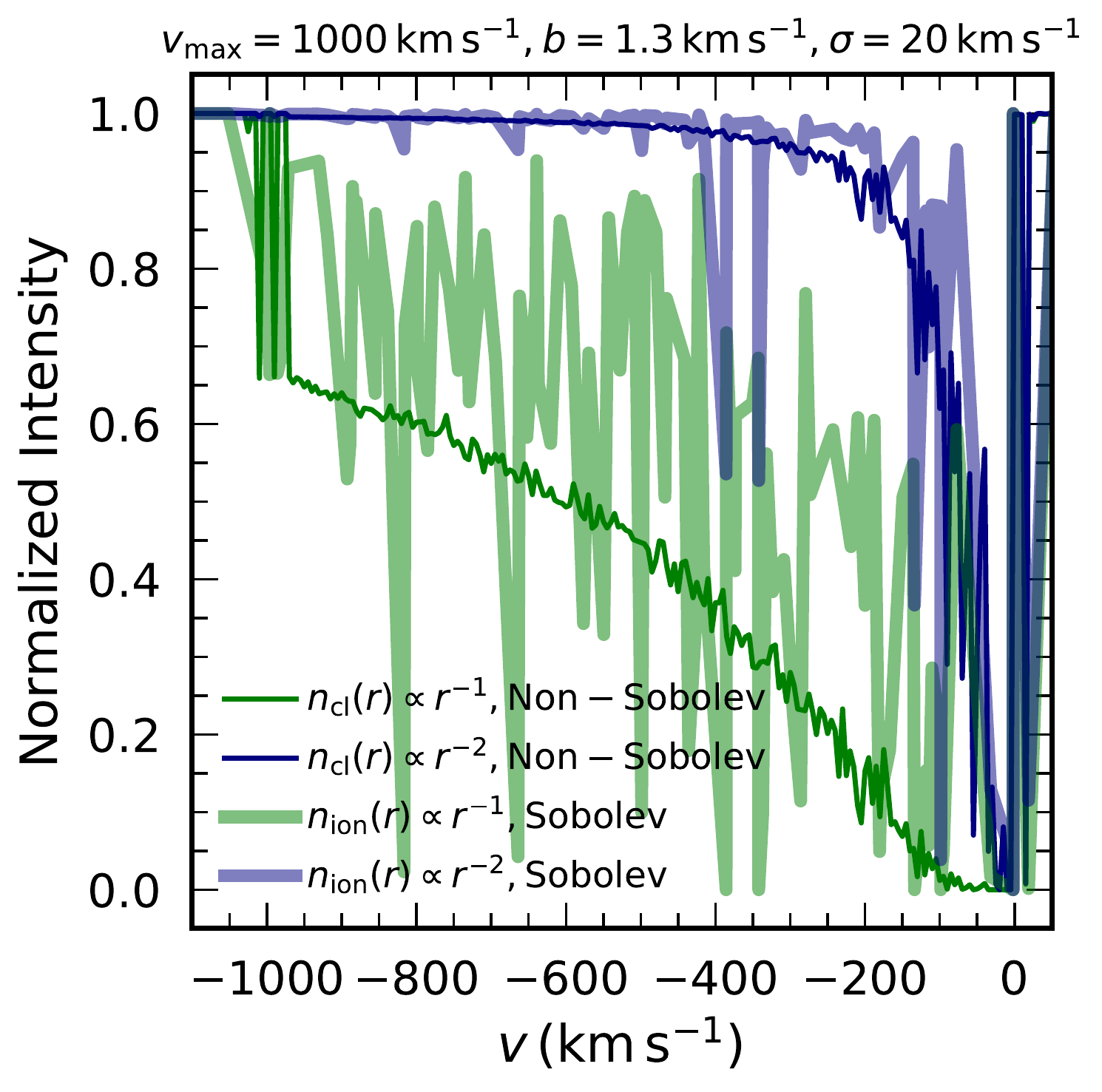}
    \caption{\textbf{Experiments designed to demonstrate the difference between the Sobolev and non-Sobolev modeling.} We consider a homogeneous medium v.s. an extremely clumpy medium where the covering fraction $C_{\rm f}(r) \simeq 1$ everywhere. \emph{Left:} If the radial velocity gradient is large, the Sobolev models (thick curves) are consistent with the corresponding non-Sobolev models (thin curves), suggesting that the Sobolev approximation works in this regime. Two sets of models with two different number density profiles of the ions or clumps ($n(r) \propto r^{-1}\ {\rm and}\ r^{-2}$) are shown in green and blue, respectively. \emph{Middle:} If the radial velocity gradient is small, the Sobolev models underpredict the amount of absorption on both the red and blue sides of the line center, suggesting that the Sobolev approximation starts to break down. \emph{Right:} If the radial velocity gradient is large but there is a small random velocity ($\sigma = 20 \kms$), the absorption line profiles predicted by the Sobolev approximation quickly become chaotic and noisy due to the stochasticity added to the velocity gradient, whereas the profiles predicted by non-Sobolev modeling remain basically unperturbed and stable. } 
    \label{fig:sobolev_non_homo}
\end{figure*}

\section{Sobolev v.s. Non-Sobolev}\label{sec:non_sobolev}

In this section, we explore the effect of the Sobolev approximation in the context of a clumpy galactic environment. The idea of the Sobolev approximation is that if the width of the absorption cross section in velocity space is much smaller than the change in the velocity of the absorbing gas within a short distance, the absorption at each velocity can be approximated as that absorption that only happens at the resonance point. Traditionally, the Sobolev approximation is usually applied to a continuous medium, such as a homogeneous wind (e.g. \citealt{Prochaska11, Scarlata15, Carr22a}). We hereby examine the use of the Sobolev approximation in a clumpy medium, and present a quantitative comparison between Sobolev and non-Sobolev modeling.

\subsection{A Homogeneous Medium v.s. An Extremely Clumpy Medium}\label{sec:homogeneous}

In a homogeneous expanding wind, the Sobolev approximation gives the line optical depth at a given radius $\tau_{\rm S}(r)$, which is solely determined by the gas number density and velocity gradient at that radius \citep{Sobolev60, Lamers99}:

\begin{equation}
\tau_{\rm S}(r) = \frac{\pi e^2
  }{m_ec} f_{\rm line} \lambda_{\rm line} n_{l}(r) \Big{|}\frac{\mathrm{d}v}{\mathrm{d}r}\Big{|}^{-1}_{r}
\label{eq:tau_s}
\end{equation}
where $n_{l}$ is the number density of the relevant ion at the lower level of the transition. The essence of Eq. (\ref{eq:tau_s}) is that it reduces the interaction between the photons and the ions to a \emph{local} process, which simplifies the calculation of optical depth (which is generally an integral over distance) to an evaluation of the properties of the absorbing gas at a single point. The DTB absorption line profile can be calculated as: 

\begin{equation}
I(v) = e^{-\tau_{\rm S}(r(v))}
\label{eq:Iv_tau_s}
\end{equation}
where $r(v)$ is the relation between velocity and radius that expresses the optical depth (and hence the line intensity) as a function of velocity.

We compare this solution obtained for a homogeneous medium with a hypothetical extremely clumpy model, where the halo is fully filled with clumps. We calculate the absorption of this model in a non-Sobolev way, meaning that the clumps are separated into a series of concentric shells (as we did in Section \ref{sec:formalism}) that all contribute to the absorption at a particular observed velocity, regardless of whether the absorption is resonant. To ensure a direct comparison with the corresponding homogeneous model, we assign the following column densities to a particular shell whose midplane is located at $r$:

\begin{equation}
N_{\rm ion,\,cl} = n_l(r) d_{\rm shell}
\label{eq:N_shell}
\end{equation}
where $n_l(r)$ is the corresponding ion number density in the homogeneous medium in Eq. (\ref{eq:tau_s}). As in an extremely clumpy medium, $C_{\rm f}(r) \simeq 1$ everywhere, the escape probability of a photon observed at velocity $-v$ is simply given by:

\begin{equation}
P_{\rm esc}(-v) = \prod_{i=1}^{N_{\rm shell}} e^{-\tau_{\rm ion} (v - v_i)}
\label{eq:pesc_cf1}
\end{equation}

Eqs. (\ref{eq:N_shell}) and (\ref{eq:pesc_cf1}) can be combined with Eqs. (\ref{eq:tauv}) - (\ref{eq:profile}) to derive the normalized absorption line intensity as a function of velocity, $I(v)$, for an extremely clumpy medium.

To compare Sobolev modeling in a homogeneous medium with non-Sobolev modeling in an extremely clumpy medium more quantitatively, we have designed several numerical experiments. For the sake of simplicity, we assign both the ions in the homogeneous medium and the clumps in the clumpy medium a radial outflow velocity profile that increases linearly with $r$:

\begin{equation}
v_{\rm out} (r) = \frac{r - r_{\rm min}}{r_{\rm h} - r_{\rm min}}v_{\rm max} 
\label{eq:linear_v}
\end{equation}
where $v_{\rm max}$ is the maximum outflow velocity achieved at $r_{\rm h}$. In this case, the radial velocity gradient, {$\mathrm{d}v_{\rm out}$} / {$\mathrm{d}r$} = $v_{\rm max} / ({r_{\rm h} - r_{\rm min}})$ is constant. For the fiducial model, we assume $n_l(r) = n_{l,\,0}(r/r_{\rm min})^{-\gamma}$ and $n_{l,\,0} = 10^{-7}\,\rm cm^{-3}$.

We present the results in Figure \ref{fig:sobolev_non_homo}. In each panel, two sets of models with $\gamma = 1.0$ and 2.0 are shown. In the left panel, we set $v_{\rm max} = 1000 \kms$ and $b_{\rm D} = 1.3 \kms$. Such a choice satisfies the ``large velocity gradient'' criterion derived by \citet{Carr22a}:

\begin{equation}
\frac{\eta}{\gamma} \gg \frac{b_{\rm D}}{v_{\rm out}(r)}
\label{eq:large_gradient}
\end{equation}
where $\eta$ is the power-law index in the velocity scaling with $r$: $v_{\rm out} \propto r^{\eta}$. For an outflow velocity profile that increases linearly with $r$, we have $\eta = 1$. In this case, the Sobolev and non-Sobolev models are fully consistent with each other, suggesting that the Sobolev approximation is working well. 

However, in the middle panel, we show the models with a decreased $v_{\rm max}$ of $100 \kms$ and an increased $b_{\rm D}$ of $13 \kms$, which corresponds to a low velocity gradient scenario that does not satisfy Eq. (\ref{eq:large_gradient}) anymore. In this case, the Sobolev models underpredict the amount of absorption on both the red and blue sides of the line center, suggesting that the Sobolev approximation is starting to break down.

In the right panel, we keep the large velocity gradient by setting $v_{\rm max} = 1000 \kms$ and $b_{\rm D} = 1.3 \kms$ but add a small random velocity ($\sigma = 20 \kms$) to the ions and clumps. In this case, the absorption line profiles predicted by the Sobolev approximation quickly become chaotic and noisy due to the stochasticity added to the velocity gradient $\mathrm{d}v / \mathrm{d}r$. In contrast, the profiles predicted by non-Sobolev modeling remain basically unperturbed and stable. In short, these experiments have demonstrated that in a homogeneous medium or an extremely clumpy medium, the Sobolev approximation only works where the velocity gradient is sufficiently large and the random motion of the gas is negligible. 

\subsection{A Not-So-Clumpy ($F_{\rm V} \ll 1$) Medium}\label{sec:not_so_clumpy}

In a realistic CGM, the clumps are likely to be non-volume-filling and hence there will be many holes in the medium that the photons can pass through freely. In this case, the Sobolev optical depth given by Eq. (\ref{eq:tau_s}) can no longer be used directly, as the absorption now depends on both the clump optical depth and the clump covering fraction. In fact, unlike the ion number density $n_{l}(r)$ or the velocity gradient $\mathrm{d}v$ / $\mathrm{d}r$, the clump covering fraction $C_{\rm f}(r)$ is not simply a function of the local properties of the clumps at $r$. In order to calculate $C_{\rm f}(r)$, one need to know the number of clumps that contribute to the geometric coverage of a sphere with radius $r$, which requires specifying and integrating over a finite width for such a sphere. This is why it is necessary to use a series of shells to properly calculate the absorption in \texttt{ALPACA}; one can also see from Eq. (\ref{eq:fc}) that $C_{\rm f}(r)$ not only depends on $n_{\rm cl}(r)$ and $R_{\rm cl}(r)$, but also the shell width $d$\footnote{Strictly speaking, one can derive a $C_{\rm f}(r)$ profile that is merely a function of $r$ and independent of the shell width $d$. For example, \citet{Dijkstra12} set $d = 2 R_{\rm cl}(r)$ and derived $C_{\rm f}(r) \simeq \frac{4}{3} \pi n_{\rm cl}(r) R_{\rm cl}^{3}(r)$, which is essentially accounting for all the clumps that can possibly intersect the sphere at $r$. In \texttt{ALPACA}, however, the shell width is always chosen to be smaller than $R_{\rm cl}(r)$, so that each clump may intersect multiple shells. In this case, we need to use the shell-width-dependent version of $C_{\rm f}(r)$ (given by Eq. \ref{eq:fc}) to avoid multiple-counting the contribution of the clumps.}.

\begin{figure*}
\centering
\includegraphics[width=0.32\textwidth]{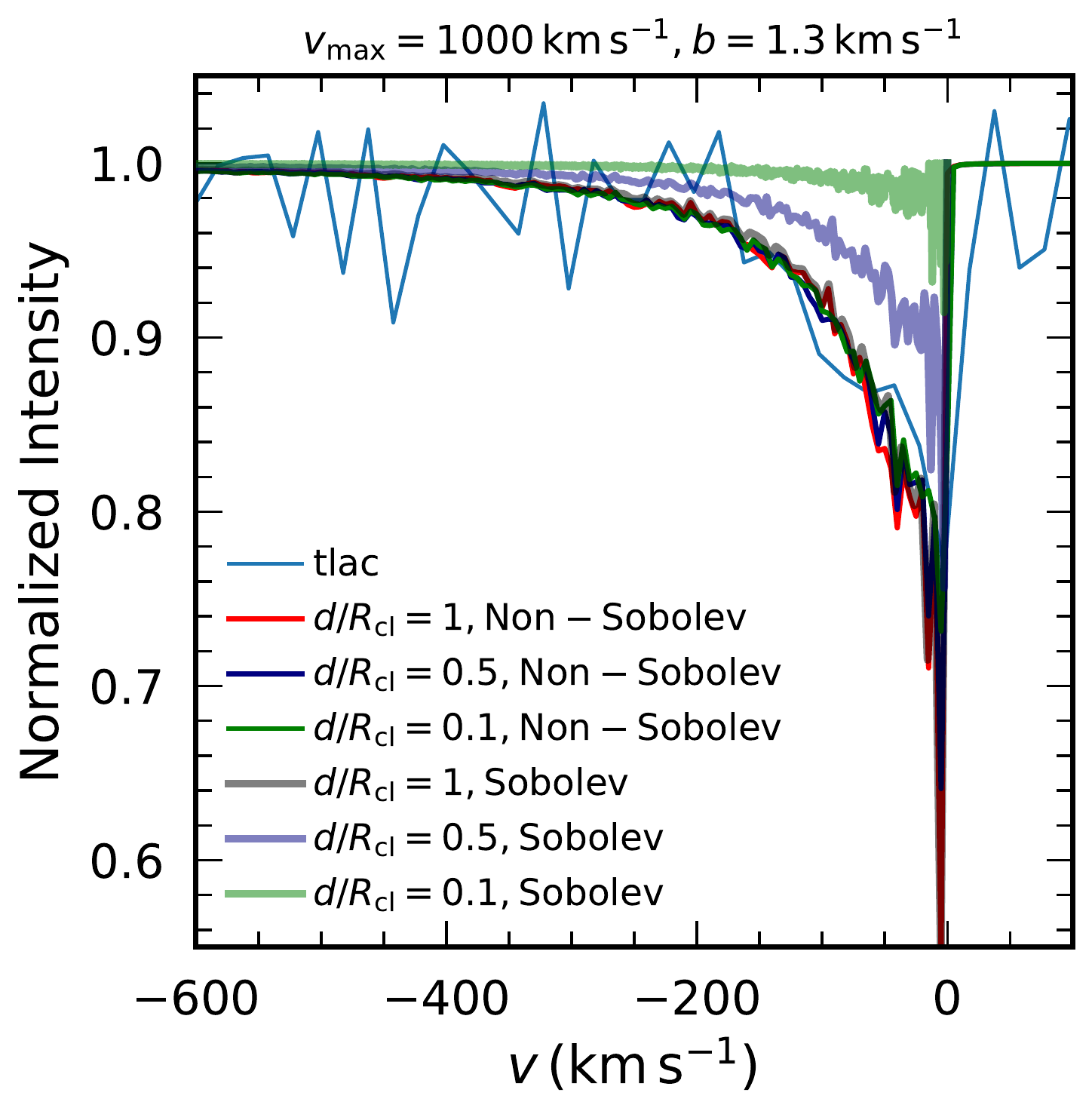}
\includegraphics[width=0.334\textwidth]{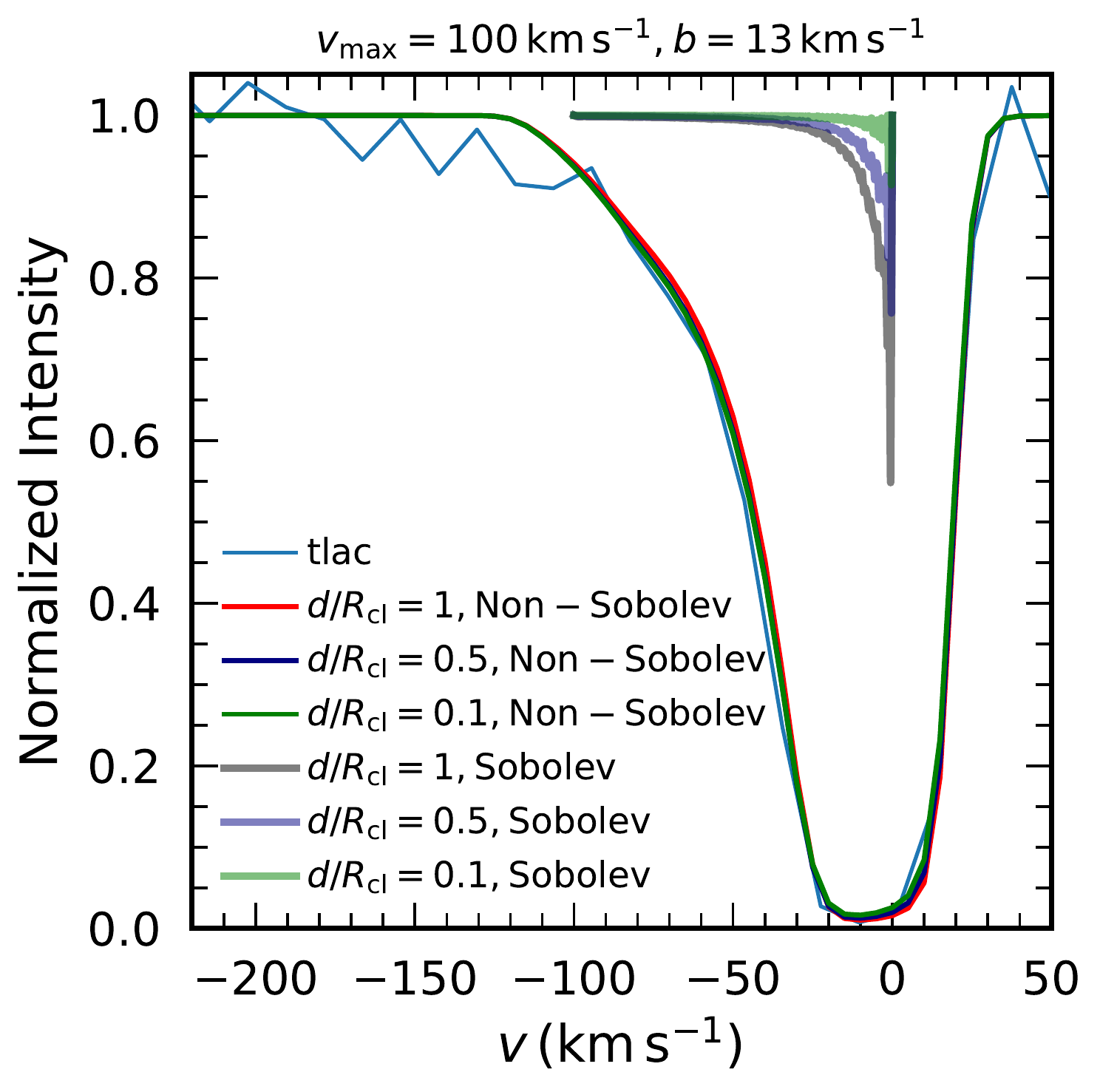}
\includegraphics[width=0.33\textwidth]{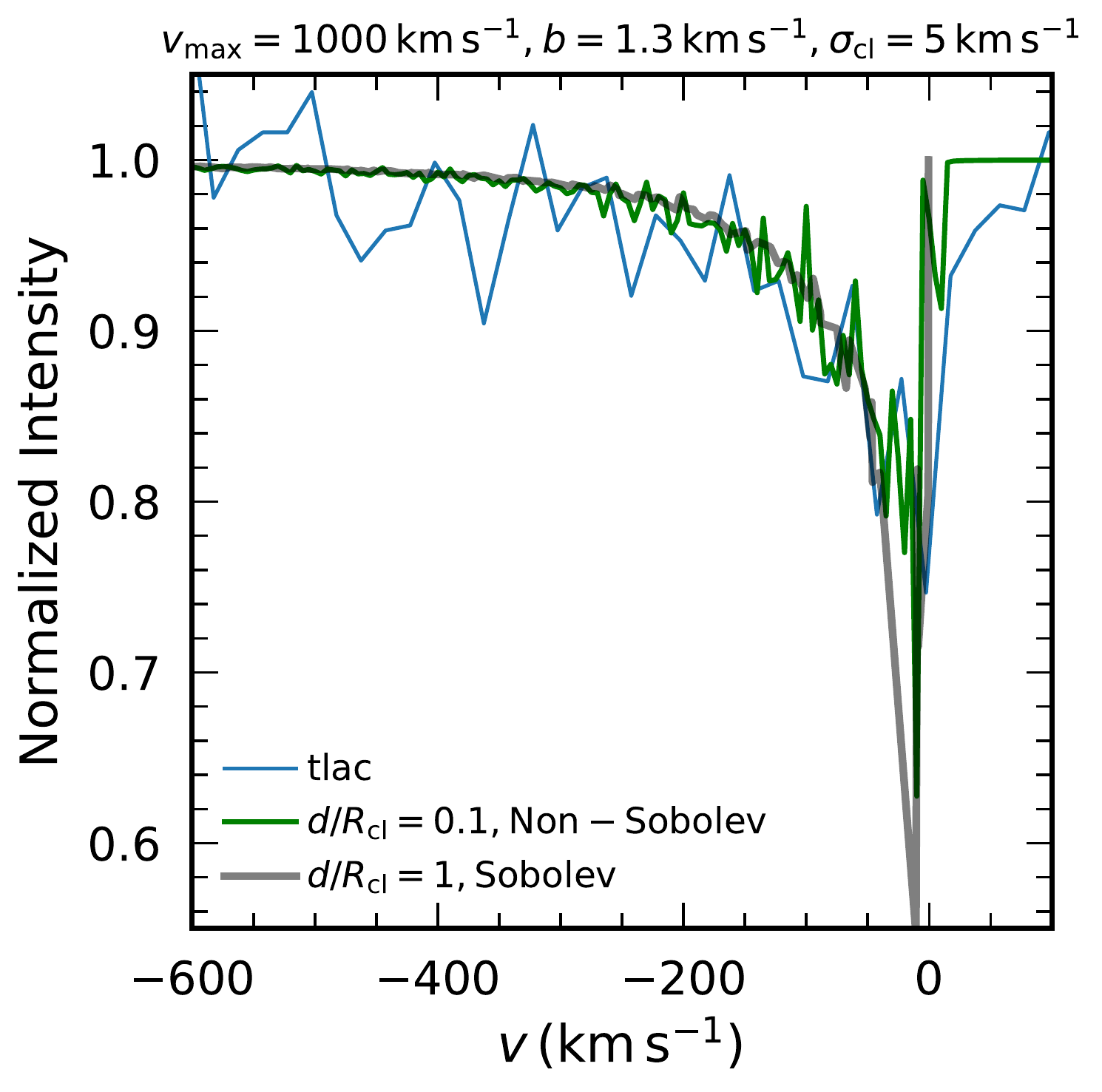}
    \caption{\textbf{Same as Figure \ref{fig:sobolev_non_homo}, but for a non-volume-filling clumpy medium.} We consider a non-volume-filling clumpy medium that contains holes through which the photons can pass freely. \emph{Left:} If the radial velocity gradient is large, the non-Sobolev models tend to converge to the Monte-Carlo simulations by \texttt{tlac} as the number of shells increases (or equivalently, as $d / R_{\rm cl}$ decreases). However, the amount of absorption predicted by the Sobolev models (shown by thick curves) decreases as the number of clumps that produce resonant absorption at each velocity has decreased. Note that the $d / R_{\rm cl} = 1$ Sobolev model is coincidentally consistent with the non-Sobolev models (see discussion in Section \ref{sec:not_so_clumpy}). \emph{Middle:} If the radial velocity gradient is small, the Sobolev models always underestimate the amount of the absorption, regardless of the choice of the number of shells. \emph{Right:} If the radial velocity gradient is large but there is a small random velocity ($\sigma_{\rm cl} = 5 \kms$),  the Sobolev model exhibits a significant deviation from the non-Sobolev model and the \texttt{tlac} prediction, suggesting that the Sobolev approximation is breaking down.} 
    \label{fig:sobolev_non_clumpy}
\end{figure*}

However, it is still possible to utilize the idea of the Sobolev approximation in a non-volume-filling clumpy medium. One can imagine that if the radial velocity gradient of the clumps is sufficiently large, the clumps that are moving at non-resonant velocities are shifted far away from resonance and will contribute negligibly to the absorption. In this case, we can possibly only account for resonant absorption and neglect all the non-resonant absorption, similar to how we applied the Sobolev approximation in a homogeneous medium. With regard to this, we similarly design several experiments to compare Sobolev and non-Sobolev modeling in such a non-volume-filling clumpy medium. In this regime, Monte-Carlo simulations from \texttt{tlac} can be conveniently performed at low computational costs to compare with the analytic models as an independent check.

We use the following set of parameters for the fiducial model: $F_{\rm V} = 0.005, {\rm log}\,N_{\rm HI,\,cl} = 14, \gamma = 2.0, {\rm and}\ R_{\rm cl} = 500{\rm\,pc}$. We then assume the linearly-increasing radial velocity profile as we do in Section \ref{sec:homogeneous}. The non-Sobolev models are generated under the standard formalism of \texttt{ALPACA}, whereas the Sobolev models are generated with the same clump parameters, but we assume that the clumps only produce resonant absorption at one particular velocity. We present the results in Figure \ref{fig:sobolev_non_clumpy}. In the left panel, we set $v_{\rm max} = 1000 \kms$ and $b_{\rm D} = 1.3 \kms$, i.e. a large velocity gradient for the clumps. We show that the model absorption line profiles predicted by \texttt{ALPACA} with an increasing number of shells (hence decreasing $d/R_{\rm cl}$ values) converge to the Monte-Carlo simulation result from \texttt{tlac}. This is because as the number of shells used in the model increases, the decrease of $C_{\rm f}(r)$ in each shell is compensated for by the increase of the total number of shells. However, the amount of absorption predicted by the Sobolev approximation (shown by thick curves) decreases as the number of shells increases, which is simply because the number of clumps that provide resonant absorption at each velocity has decreased. In other words, the model will not converge under the Sobolev approximation. Interestingly, the Sobolev model with $d/R_{\rm cl} = 1$ is actually consistent with the non-Sobolev models. We find that in this case, the velocity difference between two adjacent shells $\Delta v_{\rm cl} = v_{\rm max} / N_{\rm shell}$ is about four times as large as the clump Doppler parameter $b_{\rm D}$, which guarantees that at each observed velocity, only one shell can possibly contribute to the absorption and the other shells are all shifted far away from resonance. Therefore, in this case, whether or not using the Sobolev approximation gives the same absorption line profile.

Such a coincidental consistency between the Sobolev and non-Sobolev models is not always achievable. In the middle panel of Figure \ref{fig:sobolev_non_clumpy}, we consider another scenario where $v_{\rm max} = 100 \kms$ and $b_{\rm D} = 13 \kms$, i.e. a low velocity gradient scenario. We find that in this case, the Sobolev models always underestimate the amount of the absorption, regardless of the choice of the number of shells. This is because now $\Delta v_{\rm cl}$ cannot be much larger than $b_{\rm D}$, so that at each observed velocity, there are always multiple shells that contribute to the absorption. Only accounting for resonant absorption and ignoring the other that happens near the line center will inevitably miss a significant portion of the absorption. 

Lastly, in the right panel of Figure \ref{fig:sobolev_non_clumpy}, we set $v_{\rm max} = 1000 \kms$ and $b_{\rm D} = 1.3 \kms$ while adding a small random motion to the clumps, $\sigma_{\rm cl} = 5 \kms$. This time we only compare two models with the \texttt{tlac} simulation: a non-Sobolev model with $d/R_{\rm cl} = 0.1$ and a Sobolev model with $d/R_{\rm cl} = 1$ -- the one where the Sobolev approximation happens to predict the same line profile with the non-Sobolev models in the $\sigma_{\rm cl} = 0$ case. It can be seen that the non-Sobolev model is still consistent with the \texttt{tlac} simulations, but the Sobolev model exhibits a significant deviation from the other two models, suggesting that the Sobolev approximation is breaking down.

For the simple linear velocity profile given by Eq. (\ref{eq:linear_v}), one can estimate the condition under which the Sobolev approximation may happen to predict the correct line profile in a non-volume-filling clumpy medium. The condition that is required for the Sobolev approximation to work is:

\begin{equation}
\Delta v_{\rm cl} = v_{\rm max} / N_{\rm shell} \gg b_{\rm D}
\label{eq:Sobolev_condition}
\end{equation}
where $N_{\rm shell}$ is the number of shells used in the model. Note that the shell width cannot be chosen to be arbitrarily large; considering the clumps that can possibly contribute to the covering fraction of a shell at $r$ should have their centers located within $r\pm R_{\rm cl}$, $d/R_{\rm cl} \lesssim 2$ should be satisfied. This condition translates to the following inequality:

\begin{equation}
N_{\rm shell} \gtrsim \frac{r_{\rm h} - r_{\rm min}}{2R_{\rm cl}}
\label{eq:Nshell}
\end{equation}

Combining Eqs. (\ref{eq:Sobolev_condition}) and (\ref{eq:Nshell}), we have:

\begin{equation}
v_{\rm max} \gg \frac{r_{\rm h} - r_{\rm min}}{2R_{\rm cl}} b_{\rm D}
\label{eq:vmax}
\end{equation}

We stress that in a realistic galactic environment (e.g. CGM), this requirement is especially difficult to satisfy, considering the following two constraints: (1) the clump Doppler parameter $b_{\rm D}$ can be $10 \sim 20 \kms$ due to moderate internal turbulence (e.g. \citealt{Rudie19, Qu22}), or even $\sim 100\kms$ if a ``clump'' is actually an ensemble of smaller droplets entrained by the hot medium (e.g. \citealt{Gronke22}); (2) the clump size is often much smaller than the halo size (by two to three orders of magnitude, see e.g. \citealt{McCourt18, Zahedy19}). Moreover, note that Eq. (\ref{eq:vmax}) is derived under the assumption that the clumps' outflow velocity increases linearly with $r$. In reality, the clumps are usually accelerated to several hundred \kms\ at first, but the deceleration forces (e.g. due to gravity) start to dominate at large radii so that the velocity gradient of the clumps starts to decrease significantly. In addition, the velocity dispersion $\sigma_{\rm cl}$ of the clumps will effectively smooth out the clump velocity gradient. Therefore, the $\Delta v_{\rm cl} \gg b$ condition is at best only satisfied within a narrow range of radii where the clumps are undergoing rapid acceleration and the Sobolev approximation may be applicable. In general, applying the Sobolev approximation to the entire halo will likely result in underestimating the amount of absorption significantly. When modeling the observed absorption line profiles, any attempt to only account for ``local'' absorption will generally overestimate the clump covering fraction, as the omission of non-resonant absorption needs to be compensated for by larger clump covering fractions at different radii.

\section{Previous Work Modeling UV Absorption Lines}\label{sec:previous_works}

In this section, we briefly summarize previous work modeling UV absorption lines and compare it with \texttt{ALPACA}.

\subsection{The Picket-Fence Model}\label{sec:picket_fence}

One of the most widely used models for decoding UV absorption lines is the ``picket-fence'' model (e.g., \citealp{Steidel10, Heckman11, Zackrisson13, Jones13,  Borthakur14, Alexandroff15, Rivera15, Erb15, Vasei16, Reddy16, Rivera2017, Steidel18, Gazagnes18, Gazagnes20}), which assumes that the emitting source is partially covered by optically thick, clumpy gas material. Specifically, depending on whether dust is assumed to be present in the uncovered region (i.e. the ``holes''), the normalized line intensity can be expressed as:

\begin{equation}
I_{\lambda} = 10^{-0.4k_{\lambda}E(B-V)}(C_f e^{-\tau_{\lambda}} + 1 - C_f)
\label{eq:uniform_dust}
\end{equation}
with a uniform foreground dust screen, or: 

\begin{equation}
I_{\lambda} = 10^{-0.4k_{\lambda}E(B-V)}C_f e^{-\tau_{\lambda}} + (1 - C_f)
\label{eq:nonuniform_dust}
\end{equation}
if no dust is present in the uncovered region. In both cases, the observed photons escape either after being attenuated by the optically-thick absorbing gas (the $C_f e^{-\tau_{\lambda}}$ term) or from the holes (the $1 - C_f$ term), and dust will provide additional attenuation to the spectrum.

The picket-fence model adopts a rather phenomenological prescription for the absorbing gas, as it parameterizes the effective absorption at each wavelength or velocity empirically with the gas covering fraction $C_f$ and the effective optical depth $\tau_{\lambda}$ without considering the details of the interaction between the photons emitted at different frequencies and the atoms moving at different velocities. Depending on whether an individual line profile or a series of lines are fitted, the gas covering fraction $C_f$ can be determined as either a function of velocity (e.g. \citealt{Steidel10, Jones13, Rivera15, Rivera2017}) or as a wavelength-independent constant (e.g. \citealt{Reddy16, Steidel18, Gazagnes18, Gazagnes20}). In the former case, the Sobolev approximation (as explained in Section \ref{sec:wind_model} below) is generally adopted, and some work has used empirical treatments to account for the internal differential velocity structure of the absorbing gas, e.g. \citet{Steidel10} and \citet{Chisholm16} both considered an accelerated radial outflow with slightly different analytic forms.

\subsection{The Expanding Wind Model}\label{sec:wind_model}

Another major way of modeling the UV absorption lines is to assume a uniform, expanding wind of cool gas with radially-varying densities and velocities (e.g. \citealp{Prochaska11, Scarlata15, Carr18, Carr21}). This type of model accounts for the interaction between the emitted photons and moving atoms by using the Sobolev approximation (e.g. \citealt{Sobolev60, Lamers99}), which assumes that the photons emitted at a given wavelength or velocity interact only with the outflowing gas at a single point of resonance in the reference frame of the gas. More specifically, the absorbing gas outflowing at $v$ will only interact with the photons emitted at a frequency away from the line center by $\Delta \nu = -\frac{v}{c} \nu_0$. The outflowing gas will not have any effect on the photons that do not appear at resonance in the reference frame of the gas. Such an approximation holds when the radial velocity gradient of the clumps is much larger than the Doppler parameter of the absorbing gas, but may not necessarily be valid otherwise \citep{Carr22a}.

The expanding wind model can be used to predict absorption model spectra via either Monte Carlo RT simulations (e.g. \citealt{Prochaska11}) or semi-analytical calculations (e.g. \citealt{Scarlata15}). The model can also be upgraded to account for more complex gas geometries, e.g. hemispherical or bi-conical \citep{Prochaska11, Carr18, Carr21}, as well as additional gas kinematics, e.g. inflows \citep{Carr22}. However, in general, the model assumes a homogeneous absorbing medium and does not account for the existence of holes or clumps in the outflow \citep{Carr22a}.

\subsection{Comparison between \texttt{ALPACA} and Previous Models}\label{sec:comparison}

We have identified the following differences between the results derived from modeling metal absorption lines using \texttt{ALPACA} and previous models:

\begin{itemize}
    \item The gas covering fraction: in previous models, the gas covering fraction is either 1 (in a homogeneous medium, \citealt{Scarlata15}), or constant (in a bi-conical medium, \citealt{Carr21}), or a decreasing function with respect to $r$ (in a clumpy medium, \citealt{Steidel10, Chisholm16}). The maximum gas covering fraction derived from these models is usually of order-of-unity. In non-Sobolev modeling, however, the derived clump covering fractions at different radii are generally much smaller than one. The halo is still ``fully covered'' by clumps to an external observer, in the sense that on average there are a few clumps along any sightline. However, the required number of clumps at a particular radius or moving at a particular velocity is much lower.

    \item The gas volume filling factor: previous models that assume a homogeneous, expanding wind correspond to a volume-filling gaseous medium. Whereas in \texttt{ALPACA}, the inferred clump volume filling factor is much smaller than 1. In this sense, the results predicted by \texttt{ALPACA} are more consistent with the most current physical picture of the cool gas in a galactic environment (e.g. \citealt{McCourt18, Gronke18, Nelson20, Fielding22}). 
    
    \item The radial velocity of the absorbing gas: in previous models, the range of gas velocities strictly corresponds to the range of velocities where absorption is seen. This means if the gas is purely outflowing, no redshifted absorption will be seen \citep{Carr22a}. In \texttt{ALPACA}, however, we consider the superposition of clump outflow and clump velocity dispersion and account for all the non-resonant absorption, which makes it possible to have: (1) a maximum clump outflow velocity that is much smaller than the maximum absorption velocity (or the $v_{\rm 90}$ parameter used by previous literature) by several hundred \kms; (2) redshifted absorption without assuming the presence of external inflows. Moreover, in \texttt{ALPACA}, the clumps that contribute to the absorption observed at a particular velocity are not necessarily located all at the same radii; instead, they can be distributed at many different radial locations in the halo. In this sense, the gas absorption in \texttt{ALPACA} is more ``democratic'' than that in previous models. 
     
\end{itemize}

\section{Caveats \& Outlook}\label{sec:caveats_outlook}

In this section, we discuss several caveats that the readers should keep in mind when using \texttt{ALPACA}, as well as a number of possible applications of the model in the future.

\subsection{Caveats}\label{sec:caveats}
When we use \texttt{ALPACA} to predict DTB metal absorption line profiles, we only include the photons that travel radially and are not scattered by the clumps. Although along each sightline there are only a few clumps, the emergent absorption line profile represents the average frequency distribution of the photons that escape in all directions. In a real observation, the observed DTB absorption line profile represents the frequency distribution of the photons that emerge along the line of sight from a cylindrical region, whose radius is roughly the size of the ISM (i.e. a few kpc) and height is about the virial radius. One can estimate the number of clumps in such a cylindrical region:

\begin{equation}
N_{\rm cl,\,cyl} \simeq \frac{F_{\rm V}V_{\rm cyl}}{V_{\rm cl}} = \frac{F_{\rm V} \pi r_{\rm cyl}^2 r_{\rm h}}{\frac{4}{3}\pi R_{\rm cl}^3} 
\label{eq:Ncyl}
\end{equation}

Taking $F_{\rm V} = 10^{-3}$, $r_{\rm cyl}$ = 2 -- 3 kpc\footnote{Note that here we are considering the absorption against galaxies; for QSOs, $r_{\rm cyl}$ is much smaller and the number of clumps that contribute to the absorption along a sightline will be roughly $\sim f_{\rm c}(b)$ (see Eq. \ref{eq:fc_integrated}), whose value is generally $\sim$ a few in our model. Such a value is broadly consistent with recent observational measurements (e.g. \citealt{Zahedy19,Rudie19,Churchill20,Qu22}).}, $r_{\rm h}$ = 100 kpc and $R_{\rm cl}$ = 100 pc, we have $N_{\rm cl,\,cyl} \simeq$ 300 - 675. This is the estimated number of clumps that contribute to the absorption of the DTB spectrum of a galaxy. By comparing the model absorption line profiles predicted by \texttt{ALPACA} with the real DTB observations, we are essentially assuming the escape of photons is isotropic and the observed DTB profile is a representative sample of the frequency distribution of the escaped photons in all directions. Such an assumption may not be warranted if some of the galaxy's properties are known to be angular asymmetric (e.g. if it exhibits a collimated outflow). We plan to account for such asymmetries in our future work. 

In addition, in this work, we mainly explore an outflow-dominated kinematic profile for the clumps in the CGM of high-$z$ star-forming galaxies with a highly simplistic semi-analytic model. For other types of galaxies (e.g. quiescent early-type galaxies), the CGM gas may be inflow-dominated due to cosmological accretion (e.g. \citealt{Afruni19}). We stress that \texttt{ALPACA} is adaptable to various radial velocity profiles for the CGM gas and we will explore other possibilities in future works.

Lastly, our model does not account for any re-emission due to scattering via fluorescent channels. Our attempts to describe such fluorescent emission semi-analytically turn out to be unsuccessful, possibly because many photons are scattered multiple times and it is difficult to predict their behavior without running Monte-Carlo simulations. The infilling of fluorescent emission was considered to have a non-negligible effect on the absorption line profile of the resonant transition by previous work (e.g. \citealt{Prochaska11, Scarlata15}), yet more recent work finds that the effect of infilling is generally insignificant (\citealt{Mauerhofer21}; see also \citealt{Wang20} and \citealt{Hayes23}). More work needs to be done on both the theoretical and observational sides to better quantify the importance of fluorescent emission in the future.

\subsection{Outlook}\label{sec:outlook}

The semi-analytic model that we present in this work, \texttt{ALPACA}, serves as a complimentary tool to our \lya\ radiative transfer model, \texttt{tlac}. As of now, these two models can be used to infer the properties of ISM and CGM via modeling the following types of observational data, including but not limited to:

\begin{enumerate}
  \item spatially-resolved \lya\ profiles \citep{Erb22};
  \item \lya\ surface brightness v.s. $b$;
  \item \lya\ EW v.s. $b$;
  \item DTB metal absorption line profiles;
  \item metal absorption EW v.s. $b$.
\end{enumerate}

As we have already demonstrated in this work, the joint modeling of different datasets using \texttt{ALPACA} has the great potential of unveiling the intricate structure of galactic environments. When combined with \lya\ modeling, crucial properties of the cool gas in the CGM (such as its kinematics, see Section \ref{sec:alternative_vout_r}) can be determined reasonably well. Such a new methodology even has several far-reaching benefits for other fields in addition to galaxy evolution. For example, constraining the structure of the ISM and CGM of high-$z$ LyC emitters will help us understand how the ionizing photons propagate outwards and eventually contribute to cosmic reionization. In our next paper, we plan to apply our models on more statistically significant samples (e.g. KBSS and KLCS), with the aim of establishing a standard picture for the galactic environments of high-$z$ galaxies. We believe these efforts will eventually shed light on the nature of the galactic environments and the many important physical processes they participate in that are crucial to galaxy evolution. 

\section{Conclusions}\label{sec:conclusion}
In this work, we present \texttt{ALPACA}, a new, fast semi-analytic model for UV absorption lines that emerge from a clumpy galactic outflow. The main conclusions of this work are:

\begin{enumerate}

\item We present a semi-analytic formalism for metal absorption lines, where the galactic halo is dissected into a series of concentric shells and the photons' escape probability is a function of the clump covering fraction and velocity in each shell. With \texttt{ALPACA}, we predict the DTB metal absorption line profiles and the EW of absorption at different impact parameters as a function of the properties of the clumps, including the clump kinematics, the clump volume filling factor, the clump number density profile and the clump ion column densities. 

\item We compare the absorption line profiles predicted by \texttt{ALPACA} with the results obtained from a \lya\ radiative transfer code, \texttt{tlac}. Our tests show that the absorption line profiles predicted by \texttt{ALPACA} are consistent with the Monte-Carlo simulations performed by \texttt{tlac} over a wide range of parameters, suggesting the validity of the relatively simple formalism of \texttt{ALPACA}. We also present the effect of individual parameters of the clumps to the emergent absorption line profiles by varying each parameter individually.

\item We use \texttt{ALPACA} to jointly model the stacked DTB \CII\,$\lambda$1334 spectrum of a sample of $z \sim$ 3 LBGs and the EW v.s. $b$ profile of a sample of $z \sim$ 2 star-forming galaxy-galaxy pairs. The model successfully reproduced two datasets simultaneously, and the best-fit prefers a low volume filling factor ($\sim 3 \times 10^{-3}$) for the clumps. Moreover, the clumps' radial velocities are preferred to be a superposition of an outflow and a velocity dispersion; the outflow is rapidly accelerated to $v_{\rm cl,\,out} \sim 400 \kms$ and then gradually decelerated, whereas the velocity dispersion is $\sigma_{\rm cl} \sim 120 \kms$. The best-fit clump number density decreases with radius as $n_{\rm cl}(r) \propto r^{-1}$. As \texttt{ALPACA} accounts for clump random motion and non-resonant absorption, the best-fit model corresponds to a physical scenario where the absorption observed at a particular velocity is contributed by the clumps from a fairly broad range of radii, rather than from a single point of resonance.

\item We explore the usage of the commonly adopted Sobolev approximation in the context of a clumpy galactic environment. We find that in an extremely clumpy medium that resembles a homogeneous medium, the Sobolev approximation only works when the velocity gradient is sufficiently large and the random motion of the gas is negligible. Whereas in a realistic, non-volume-filling clumpy medium, the Sobolev approximation is at best only applicable within a narrow range of radii where the clumps are undergoing rapid acceleration and fails otherwise. Applying the Sobolev approximation to the entire halo of a galaxy has the risk of overestimating the clump covering fraction significantly.

\item We find that the clump radial velocity profile may not be fully constrained by the joint modeling of the DTB spectrum and the EW v.s. $b$ profile. The analysis of additional observational data, such as spatially-resolved \lya\ emission modeling, may help break the degeneracy and distinguish different clump radial velocity profiles.

\end{enumerate}

\section*{Data availability}
The data underlying this article will be shared on reasonable request to the corresponding author. 

\section*{Acknowledgements}

We thank Phil Hopkins for providing computational resources. ZL acknowledges Cody Carr for the illuminating discussion. MG thanks the Max Planck Society for support through the Max Planck Research Group. CS and ZL have been supported in part by grant AST-2009278 from the U.S. National Science Foundation. Numerical calculations were run on the Caltech compute cluster ``Wheeler,'' allocations from XSEDE TG-AST130039 and PRAC NSF.1713353 supported by the NSF, and NASA HEC SMD-16-7592. We also acknowledge the use of the the following software packages: Astropy \citep{Astropy18}, the SciPy and NumPy system \citep{Scipy20, Numpy20}.

\bibliographystyle{mnras}
\bibliography{Clumpy}

\appendix

\section{Model Convergence}\label{sec:convergence}

\begin{figure}
\centering
\includegraphics[width=0.47\textwidth]{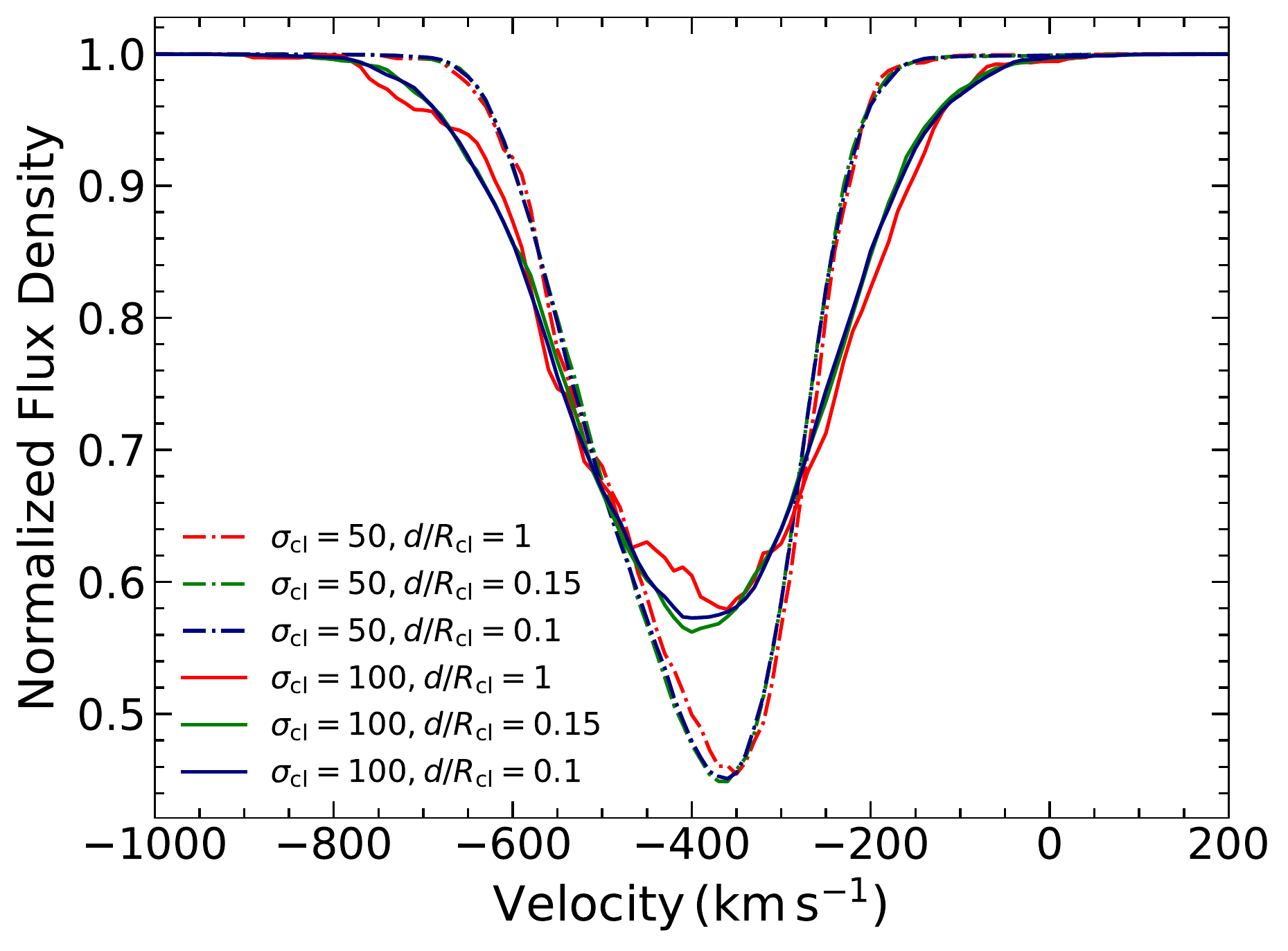}
    \caption{\textbf{Testing the convergence of \texttt{ALPACA} by varying $d / R_{\rm cl}$. } The fiducial model parameters are given in the text, and two sets of models with $\sigma_{\rm cl} = 50\ {\rm and}\ 100 \kms$ are calculated (shown in dash-dotted and solid lines, respectively). In each model set, the number of shells is varied (hence the $d / R_{\rm cl}$ values) and the model spectra are shown in different colors. For both sets of models, the convergence is achieved at $d / R_{\rm cl} \sim 0.1$.
    \label{fig:convergence}}
\end{figure}

In this section, we test how many shells are needed to achieve convergence for \texttt{ALPACA}. We choose the following set of parameters as our fiducial model: $F_{\rm V} = 0.005, \mathcal{V} = 700 \kms, \alpha = 2.0, \gamma = 0.0, {\rm log}\,N_{\rm HI,\,cl} = 15, R_{\rm cl} = 100{\rm\,pc}, \sigma_{\rm cl} = 50 \kms, b_{\rm D} = 15 \kms$. We find that changing any of the parameters does not visibly affect the condition for convergence, except for $\sigma_{\rm cl}$, which appears to have a moderate effect on the model spectra with different number of shells. Therefore, we test the model convergence by varying $d / R_{\rm cl}$ ($d = {(r_{\rm h} - r_{\rm min})}/{N_{\rm shell}}$ is the spacing between the midplanes of two adjacent shells, where $N_{\rm shell}$ is the number of shells) with $\sigma_{\rm cl} = 50\ {\rm and}\ 100 \kms$. As shown in Figure \ref{fig:convergence}, in both sets of models, the variation in the model spectra starts to become negligible at $d / R_{\rm cl} \sim 0.1$, suggesting the convergence of the model has been achieved. We have also verified increasing $N_{\rm shell}$ even more no longer visibly changes the model absorption line profile.

\bsp	
\label{lastpage}
\end{document}